\documentstyle[12pt]{article}

\newcommand{\dyy}{\begin{equation}}
\newcommand{\dzz}{\end{equation}}
\newcommand{\daa}{\begin{eqnarray}}
\newcommand{\dbb}{\end{eqnarray}}

\newcommand{\sksl}{{\mbox{\boldmath $\sigma$}}_k\cdot{\mbox{\boldmath $
            \sigma$}}_l}
\newcommand{\lkll}{\lambda_k^a\cdot\lambda_l^a}
\newcommand{\titj}{{\mbox{\boldmath $\tau$}}_i\cdot{\mbox{\boldmath$\tau$}}_j}
\newcommand{\sisj}{{\mbox{\boldmath $\sigma$}}_i\cdot{\mbox{\boldmath $
            \sigma$}}_j}
\newcommand{\lilj}{\lambda_i^a\cdot\lambda_j^a}
\newcommand{\sigi}{{\mbox{\boldmath $\sigma$}}_i}
\newcommand{\sigj}{{\mbox{\boldmath $\sigma$}}_j}
\newcommand{\asym}{ \parbox{.4cm}{
                    \begin{picture}(.3,.6)
                      \thicklines
                      \put(0.,.3){\framebox(.3,.3){{\mbox{\rm\tiny i}}}}
                      \put(0.,0.){\framebox(.3,.3){{\mbox{\rm\tiny j}}}}
                    \end{picture} }_C }
\newcommand{\sym}{  \parbox{.7cm}{
                    \begin{picture}(.6,.3)
                      \thicklines
                      \put(0.,0.){\framebox(.3,.3){{\mbox{\rm\tiny i}}}}
                      \put(.3,0.){\framebox(.3,.3){{\mbox{\rm\tiny j}}}}
                    \end{picture} }_C }
\newcommand{\sing}{ \parbox{.4cm}{
                    \begin{picture}(.3,.9)
                      \thicklines
                      \put(0.,0.){\framebox(.3,.3){}}
                      \put(0.,.3){\framebox(.3,.3){}}
                      \put(0.,.6){\framebox(.3,.3){}}
                    \end{picture} }_C }
\newcommand{\singl}{\parbox{.7cm}{
                    \begin{picture}(.6,.9)
                      \thicklines
                      \put(0.,0.){\framebox(.3,.3){}}
                      \put(0.,.3){\framebox(.3,.3){}}
                      \put(0.,.6){\framebox(.3,.3){}}
                      \put(.3,0.){\framebox(.3,.3){}}
                      \put(.3,.3){\framebox(.3,.3){}}
                      \put(.3,.6){\framebox(.3,.3){}}
                    \end{picture} }_C }

\begin{document}

\setlength{\baselineskip}{.3in}

\begin{titlepage}

\begin{center}
{\large\bf CONSTITUENT QUARK MODEL CALCULATION FOR}

{\large\bf A POSSIBLE J$^P$=0$^-$, T=0 DIBARYON}
\end{center}

\begin{center}
{\large Georg Wagner, L.Ya.\ Glozman$^\ast$, A.J.\ Buchmann, Amand Faessler}

(Institut f\"{u}r Theoretische Physik, Universit\"{a}t T\"{u}bingen,
Auf der Morgenstelle 14, D -- 72076 T\"{u}bingen, Germany)
\end{center}

\begin{center}
{\large Abstract:}
\end{center}

\medskip
\noindent
There exists experimental evidence that
a dibaryon resonance {\bf $d'$} with quantum numbers
J$^P$=0$^-$, T=0 and mass $2065$ MeV
could be the origin of the narrow peak in the
$(\pi^+ ,\pi^- )$
double charge exchange cross--sections on nuclei.
We investigate the six--quark system with these quantum--numbers within
the constituent quark model, with linear confinement,
effective one--gluon exchange at short range
and chiral interactions between quarks ($\pi$-- and $\sigma$--exchange).
We classify all possible six quark states with
J$^P$=0$^-$, T=0, and with N=1 and N=3 harmonic oscillator excitations, using
different reduction chains.
The six--quark Hamiltonian is diagonalized in the basis including
the unique N=1 state and the 10 most important states from the N=3 shell.
We find, that with most of the possible sets of parameters, the mass of
such a "dibaryon" lies above the N(939)+N$^\ast$(1535) threshold.
The only possibility to describe the supposed d'(2065) in the present
context is to reduce the
confinement strength to very small values, however at the expense of
describing the negative parity resonances N$^\ast$.
We also analyze the J$^P$=0$^-$, T=2, N=1 six--quark state.

\vspace{1.2cm}
\noindent
$^\ast$Alexander von Humboldt Fellow,
on leave from Alma--Ata Power Engineering Institute,
Kosmonavtov 126, Alma--Ata, Kazakhstan.

\end{titlepage}

\setlength{\unitlength}{1cm}
\section{Introduction}

Believing QCD to be the theory of strong interactions,
we consider hadronic systems as built up by quarks and gluons.
With the exception of lattice calculations, which start directly
from the underlying current quark and gluon fields,
the descriptions of baryons make use of some effective degrees of freedom of
nonperturbative QCD, such as constituent quarks \cite{gis}, chiral fields
\cite{wit}, etc.
In many--baryon systems, like atomic nuclei, the relevant degrees of
freedom are  baryons and mesons (except
for perhaps some special cases \cite{gno}, where the quark degrees of
freedom must be taken into account explicitly).

However, in principle nothing forbids the existence of objects
with baryon number bigger than one, which have some QCD motivated origin and
could be considered as built up with quarks and gluons rather than baryons
and mesons. Such objects would have, in contrast to atomic nuclei, quite
a small size, of order of 1 fm.

This idea was very popular about 10--15 years ago and a lot of efforts
were made to find such objects, both theoretically and experimentally
(see, for example, refs.\cite{jaf}--\cite{str} and references therein).
However, an experimental search for dibaryons in the NN--system
has not been successful until now.

The reason for this seems to be quite clear. The only demand
for the color part of the six--quark wave function of a hypothetical
dibaryon is that it is a color singlet, i.e.\ it has a symmetry
\dyy
[2^3]_C \equiv\; \singl \quad .
\label{eq:sixsix}\dzz
This singlet $SU(3)_C$ representation contains in its Clebsch--Gordan
expansion the state
\dyy
[1^3]_C \;\otimes\; [1^3]_C \equiv \;\sing\otimes\;\sing
\label{eq:sixsinglet}
\dzz
(Here and in the
following, "$\times$" denotes an inner product, whereas
"$\otimes$" denotes an outer one).
But in this case the strong confinement forces
$\sim \lilj$
are absent between two color singlet objects:
With the usual normalization
of the Gell--Mann matrices $\vec\lambda^2 \equiv
\lambda^a\lambda^a = \frac{16}{3}$ (summation over double indices), one finds
\dyy
\langle\;\sing\otimes\sing\vert\, \lilj \,\vert
\;\sing\otimes\sing\rangle = 0 \qquad , \qquad \left\{ \begin{array}{l}
i\subset\{ 1,2,3\}  \\
j\subset\{ 4,5,6\} \end{array} \right. \quad .
\label{eq:colorforce}
\dzz

So, if two baryons are very close to
each other  and form a 6q--system, the
confinement forces will not prevent such a system from a decay
into baryon--baryon channels. This means, that the lifetime for such a
system would be extremely small and the corresponding large width would thus
not allow to treat such a system as a dibaryon. The only possibility for
the dibaryon in this case would be if there were some forces
(like $\lilj\sisj \; ,\; \titj\sisj$,  etc.)
which bind the six--quark system
below the corresponding baryon--baryon
threshold for a strong decay.
This argument stimulated for example the search for
an H--particle \cite{jaf,ot,str}.

\smallskip
On the other hand, there seems to be some
recent experimental evidence for a negative parity resonance in the
$\pi NN$--system
with the quantum numbers J$^P$=0$^-$, T=0 \cite{bcs}, for which the notation
d' has been introduced.
The pionic double charge exchange
reaction (DCX) on nuclei
\dyy
{^A\! Z}\; (\pi^+,\pi^-) \; {^A\! Z+2}
\label{eq:reaction}
\dzz
exhibits, independently of the nuclear target, a very narrow peak near
the pion kinetic energy
$T_\pi = 50$ MeV. Its position as well as its width are practically identical
for the available world DCX--data on light and medium
nuclei such as $^{12}$C, $^{14}$C, $^{18}$O,
$^{44}$Ca, $^{48}$Ca. Only its amplitude depends on the considered nucleus.
This suggests, that it can only be connected to some elementary process.

Due to charge conservation this elementary
process  involves at least two
nucleons within the nucleus. The hypothesis, that this peak is
due to a dibaryon resonance $d'$ in the $\pi NN$ channel
with quantum
numbers J$^P$=0$^-$, T=0 and mass $2065$ MeV \cite{bcs}, allows to
describe all available data. This dibaryon cannot
decay into the nucleon--nucleon channel due to the Pauli principle
and its mass is below any baryon--baryon ($NN^*, N^*N^*,...$) threshold.
These peculiarities explain the very small width of only
$\Gamma\sim 5$ MeV of the d' dibaryon.

\smallskip
These circumstances motivated our detailed study of the six--quark system
with the quantum numbers
J$^P$=0$^-$, T=0 within a constituent quark model \cite{gno,oy,ffl,ok,buc}
with and without
chiral interactions between constituent quarks.

\smallskip
Dibaryons with
quantum numbers J$^P$=0$^-$, T=0 were already studied in the framework
of string--like (deformed) bag models and were calculated as
strings with $q^4$ and $q^2$ colored quark clusters at the ends
\cite{mul,kms}.
An essential shortcoming of this model is the lack of
antisymmetrization between quarks belonging to  different clusters.
Antisymmetrization can be neglected only for well separated
clusters. However, in the $q^4 - q^2$ system the color--exchange
(confining) forces between
quarks do not allow large separations, and the typical size of
such a system is expected to be of order 1 fm. We know from  experience
in the $NN$--system  that the Pauli
principle on the quark level plays a decisive role at such
distances \cite{gno,oy,ffl}.

The other drawback of the bag--model in the six--quark system is that the
boundary conditions (which simulate the quark confinement) prevent such a
system from the color--singlet 3q -- color--singlet 3q clusterization,
which is the most important phenomenon in 3n--quark ($n>1$) systems.
If such a model were correct, a large number of dibaryons should be observed.
 From this point of view, a potential model, where quark confinement is
approximated by two--body q--q forces $\sim\lilj$, is more satisfactory,
although one has to take care
of the long--range Van--der--Waals forces in this case,
when describing the NN--system \cite{gl}.

\smallskip
In this work we study the six--quark system with the
quantum numbers J$^P$=0$^-$, T=0 within the constituent quark
model \cite{gno,oy,ffl,ok,buc}
taking into account properly the Pauli principle throughout the whole
calculation.
We investigate both possibilities; with a chiral field (pions and sigma mesons)
coupled directly to
constituent quarks, and without such a chiral field. In the latter case,
quarks interact only through color--exchange potentials. In our previous
communication \cite{gbf}, only the lowest shell--model state $s^5p$ was used.
In that paper, we did not care about a correct description of the
one--quantum nucleon excitations N$^\ast$(1535) (J$^P$=$\frac{1}{2}^-$,
T=$\frac{1}{2}$). However, this is quite important, since the dibaryon mass
should be compared with the N(939)+N$^\ast$(1535) threshold.
Here we present a more complete
calculation taking into account configuration mixing,
including the 10 presumably
most important excited states.
We find that with all sets of parameters, that describe rather well
the lowest baryon mass spectrum, the mass of the
J$^P$=0$^-$, T=0
6q--state lies around 100 -- 200 MeV
above the NN$^\ast$--threshold. However, including more excited states
in our basis, this mass could come down below the threshold by some 10 MeV's.
If it is so, there really could exist a dibaryon with these
quantum numbers.
On the other hand, its mass, at least within our model, is still higher
than the experimentally observed peak suggests.
Assuming that the confinement potential in a six quark system is weaker than
the usual choice in three quark systems (This would still describe the nucleon
and the $\Delta$, but not the negative parity
N$^\ast$ resonances.), yields a
dibaryon with a mass close to 2065 MeV and an oscillator length
$b_6\simeq 1.25$ fm.

It seems that this 6q--system is very sensitive to
the explicit form of the confinement mechanism, or, in other words, the
extension of the baryon confinement to multi--quark confinement might not
be too straightforward. In this sense,
the confirmation of the d'
resonance could give valuable information to our understanding of confinement.

\smallskip
The structure of the paper is the following:
The effective Hamiltonian for the dibaryon
is presented in section 2, and some short
motivation will be given for its different ingredients.
Fitting the parameters to the spectrum of light baryons
is the commonly accepted procedure
to fix the effective Hamiltonian.
In section 3, we present the Translationally Invariant Shell--Model
(TISM) basis.
We classify the basis states for the
6q--system with J$^P$=0$^-$, T=0,
and review some important
points concerning the fractional parentage technique for the
wavefunction.
We present the results for the dibaryon, first excluding the mixing
of excited states (cf.\ \cite{gbf}), and then including the ten presumably
most important excited states.
A short discussion of a possible J$^P$=0$^-$, T=2 six--quark state is
also added. Section 7 gives a conclusion of the present work.
All necessary analytical expressions are reserved for the appendices.

\section{The effective Hamiltonian}

One now generally assumes,
that the constituent quarks are quasiparticles with a
complicated structure and dynamical (q--dependent) mass
\cite{mg}--\cite{vw}. The current
quarks of the underlying QCD acquire the constituent mass due to the
spontaneous breaking of the chiral symmetry of the QCD Lagrangian, which
is nearly exact in the $SU(2)$--flavour sector.

This chiral symmetry breaking is caused probably
by the instanton structure of the
QCD vacuum \cite{dp,shu}
and is characterized by the corresponding order parameter
-- the non--zero quark condensate -- and by the Goldstone exitations -- the
pions. All these features are also reproduced in the
Nambu--Jona-Lasinio model (see recent review \cite{vw}
and references therein),
which can be considered as an effective approximation to the underlying
QCD in the low--energy domain.

So, the low--momentum Lagrangian must contain the constituent quarks and
chiral fields as effective degrees of freedom.
But it must also contain a color--exchange interaction between
the constituent quarks, since the latter are coloured objects. Among
these effective interactions we should include a confinement force
(whose real nature is not known), and also
a short--range (i.e.\ at distances less than the chiral symmetry breaking
scale) effective gluon--exchange between quarks.

\smallskip
There is a very close analogy to this picture coming from solid state
physics.
The correct effective degrees of freedom for an explanation of thermo--
and/or electric properties of metals are not light elementary electrons
and ions, but heavy electrons with effective mass $m^*$ and phonons.
Phonons are  Goldstone excitations and appear as a result
of the breaking of translational invariance in the lattice of  ions.
This spontaneous breaking is also characterized by the effective electron
mass $m^*$ (microscopically, this effective mass arises due to very
complicated interactions between the elementary electrons and the
lattice). There are also some residual interactions such as
electron--electron, electron--phonon, etc.

After these quite general remarks and motivation for our chosen quark picture,
let us now turn to the concrete description of the effective Hamiltonian.

\vspace{.5cm}
\noindent
{\bf The chiral sector: Pion-- and Sigma--exchange}

\medskip
The chiral--invariant interaction Lagrangian of the $\sigma$--model type
\cite{vw,gel,ok}, which describes
in our case an interaction of the constituent quark field
$\Psi$ (spinor in $SU(2)_{flavour}$--space) and the chiral fields
$\mbox{\boldmath $\pi$}$ and $\sigma$, is
\dyy
{\cal L}_{int} =
-g\overline{\Psi}(\sigma + i\gamma_5\mbox{\boldmath $\pi\cdot\tau$})\Psi.
\label{eq:sigmamodel}\dzz

The constituent quarks have a complicated structure which can be parametrized
by a formfactor in the quark--chiral field vertex.
For that, we use the substitution
\dyy
g\longrightarrow g\; {\left({\Lambda^2\over{\Lambda^2 +
{\bf k}^2}}\right)}^{1/2}
\label{eq:cutoff}\dzz
where $\Lambda$ characterizes the chiral symmetry spontaneous breaking scale,
and must be of the order of the instanton size in the QCD vacuum
\cite{shu}, i.e.\ $\Lambda\simeq {\rm (0.7 - 1.5) GeV.}$

This cut--off contributes to the pion--nucleon interaction radius as
\dyy
\langle r^2\rangle_{\pi N} = b^2 +\frac{3}{\Lambda^2} \qquad ,
\label{eq:pinuc}\dzz
where b is the
quark core size of the nucleon \cite{buc}.
The larger the cut--off $\Lambda$, the
more pointlike behaves the pion--quark coupling. Intuitively, one could think
of this cut--off as an effective size of the constituent quarks
$\langle r^2_q\rangle = \frac{3}{\Lambda^2}$.
For example $\Lambda$=4.2 fm$^{-1}$ would correspond to a size of
$\sqrt{\langle r_q^2\rangle} \simeq$ 0.4 fm.
On the other hand, the cut--off formfactor behaviour in eq.(\ref{eq:cutoff})
can be related to the $q^2$--dependence of the
dynamical constituent quark mass
\cite{ok}.

The sigma mass $m_{\sigma}$ is fixed by the relation $m_{\sigma} \simeq
2m_q,$  where $m_q \simeq \frac{1}{3} m_N$ is the constituent
quark mass. On the other side, the relation
$$m_\sigma^2 = 4m_q^2 + m_\pi^2 $$
appears as a result of the bosonization of the Nambu and Jona-Lasinio
Lagrangian.

The pion--quark coupling constant $g$ is determined by the well
known $g_{\pi N}$ coupling constant $g = \frac {3}{5} \frac
{m_q}{m_N}\, g_{\pi N}$ ($g_{\pi N}=13.36$), where the coefficient
$\frac{3}{5}$
comes from the spin--isospin matrix element when we consider the
$\pi N$ interaction as the interaction between the pion and the
3 constituent quarks.

Our notations for the coupling constants are
\dyy
f_{\pi q} =\frac{m_\pi}{2m_q}\cdot g\quad ; \quad
g_{\sigma q} =\frac{m_\sigma}{2m_q}\cdot g\simeq g\quad .
\label{eq:coupconst}\dzz

\medskip
Fourier--transforming the static pion propagator,
\dyy
\frac{1}{(2\pi )^3}\int \exp (i\vec q\cdot\vec r) \;\frac{d^3\vec q}{\mu^2 +
\vec q^2}\quad =\quad \frac{1}{4\pi} \;  \frac{\exp (-\mu r)}{r}
\label{eq:staticpion}\dzz
we arrive at the usual Yukawa--like potential for the isovector pion
(and similarily for the scalar--isoscalar sigma, which is not sensitive to
any spin--isospin quantumnumbers of the quarks, but only to their orbital
distribution) :
\daa
V^{\bf\pi}_{ij}({\bf r})\!\!\! &=& \!\!\! {\Lambda^2\over\Lambda^2-m_\pi^2}
{f_{\pi q}^2
\over 4\pi}
\frac{\titj}{3} \bigg(\! \big(\sisj
V_C(m_\pi r)
\! +\!\hat S_{ij} V_T(m_\pi r) \big) -{\Lambda^2\over m_\pi^2}
(m_\pi\!\rightarrow\!\Lambda )\!\bigg) \nonumber\\
& & \label{eq:opep}\\
V^{\bf\sigma}_{ij}(r)\!\!\! &=&\!\!\! -{\Lambda^2\over\Lambda^2-m_\sigma^2}\,
{g^2\over 4\pi}
\bigg( V_C(m_\sigma r) - V_C(\Lambda r) \bigg)\, ,\label{eq:osep}\\
V_C(mr)\!\!\! &=&\!\!\! {\exp (-m r)\over r}\qquad ;\qquad
V_T(mr) = \bigg( 1+{3\over mr}+{3\over m^2r^2}\bigg)
{\exp (-m r)\over r} \, ; \label{eq:yukawa}\\
\hat S_{ij}\!\!\! &=&\!\!\! \bigg( {3\sigi\cdot{\bf r}\;\;\sigj\cdot{\bf r}
\over r^2} - \sigi\,\cdot\,\sigj \bigg)
\label{eq:tensorop}\dbb

\vspace{.5cm}
\noindent
{\bf The Color--part: Confinement and Gluon exchange}

\medskip
We take  the confinement potential, which mainly determines the
medium-- and long--range (on the quark scale) phenomena, to be linear
\dyy
V^{Conf}_{ij}(r\!=\!\vert {\bf r_i} - {\bf r_j}\vert )
= -\; {\lambda_i^a \cdot\lambda_j^a \over 4} \; \big( a_c r - C\big)
\label{eq:conf}\dzz

The effective "one--gluon exchange" potential, which is
responsible for the very short--range phenomena and motivated in
its form from the charmonium spectroscopy \cite{rgg}, contains essentially
five terms:
\daa
V^{OGE}_{ij}({\bf r=r_i-r_j})
&=& {\alpha_s\over4} \lambda_i^a  \cdot \lambda_j^a
\Bigg\{ ~{1\over r} - {\pi\over{m_q^2}} \left( 1 + {2\over3}
\sisj \right) \delta({\bf r}) -\nonumber\\
& & -{1\over{4m_q^2r^3}} \cdot \hat S_{ij}
- {3\over{m_q^2r^3}}{\bf r}\times{\bf
p}_{ij}\cdot \left({\mbox{\boldmath $\sigma$}}_i
+ {\mbox{\boldmath $\sigma$}}_j  \right) \Bigg\}
\label{eq:ogep}\dbb
the so--called color--coulomb (CC $\simeq\lilj\;\frac{1}{r}$),
the contact color--electric (CE
$\simeq\lilj\;\delta (\vec r)$) term,
the contact color--magnetic (CM $\simeq\lilj\sisj\;\delta (\vec r)$) term,
the tensor term proportional $\hat S_{ij}$ and the
Galilei--invariant spin--orbit term.

In our present study, we drop
the spin--orbit term. As it is well known, there is practically no room
for the spin--orbit interaction when describing the baryon spectrum. It is
well seen, for example, from the small mass--splitting of the
$\frac{1}{2}^-$ and
$\frac{3}{2}^-$ resonances N$^\ast$(1535) and N$^\ast$(1520). Usually it is
assumed, that the Galilei--invariant
spin--orbit forces from the OGE--potential are
compensated in the baryons by the spin--orbit forces from other sources,
e.g.\ the Thomas term coming from the confinement \cite{gis}
 or the $\sigma$--exchange
\cite{vbf}.

\vspace{.5cm}
\noindent
{\bf Parameter fitting to light baryons}

\medskip
The constituent
quarks, $\pi$-- and $\sigma$--fields, and the color--exchange forces are
some effective degrees of freedom of  nonperturbative QCD. They do not
follow from first principles in a compelling and unique way, but are only
supposed to simulate the dominant features of QCD in the low energy domain.
The strong coupling constant $\alpha_s$ for example, is not the
momentum--dependent running coupling constant of QCD, but an effective
parameter, which has to be fixed to some observable of
low energy QCD to define our Hamiltonian.

A possible way to
fit the 5 parameters of our final Hamiltonian of eq.(\ref{eq:ham}),
$\alpha_s, a_c, C,
\Lambda$ and $m_q$, is to describe the baryon spectrum, or more precisely
the light baryons such as the nucleon, the $\Delta$--resonance, and
the J$^P$=$\frac{1}{2}^-$,T=$\frac{1}{2}$ one--quantum excitation
N$^\ast$(1535) (the reason, why one should fit our parameters first of all
to the N$^\ast$(1535), but not to the other excited states, will be seen from
considerations in section 4). For the N and $\Delta$, we use the simple
$s^3$ harmonic oscillator functions (without the center--of--mass
oscillations: see states $\vert N_0\rangle$ and $\vert\Delta_0\rangle$ in
appendix A and in
refs.\cite{gno,buc}). The analytical expressions for the mass--formulae
$m_N = h_{00}^N$ and $m_\Delta = h_{00}^\Delta$ can be found in appendix A.

To suppress the mixing with the two-- and more--quantum harmonic
oscillator states in the N and the $\Delta$, we impose the nucleon
stability condition \cite{ffl}
$\frac{\partial m_N}{\partial b_N} =0$, where $b_N$ is the
harmonic oscillator parameter in the 3--quark system (which coincides with
the mean--square quark core radius of the N and the $\Delta$).

For the N$^\ast$(1535), we diagonalize analytically our Hamiltonian in
the space including the two possible negative parity one--quantum states
$\vert N_1^{(-)}\rangle$ and
$\vert N_2^{(-)}\rangle$ in notations of the second paper in
ref.\cite{gno} (All the
necessary wave functions can be found there), and fit the smaller
eigenvalue to the N$^\ast$(1535) mass. The corresponding
mass--matrix can also be found in
appendix A\footnote{In this basis,
the 2nd eigenvalue can be compared to the next one--quantum
excitation with J$^P$=$\frac{1}{2}^-$,T=$\frac{1}{2}$, the N$^\ast$(1650).
We give this result for illustration
in table \ref{table:paraset}.}.

\smallskip
Thus, for the six free parameters,
$\alpha_s,a_c,C,\Lambda ,b_N,m_q$, we have now four constraints. But there
is not to much freedom because one has the following additional, approximate
constraints:

\noindent
i) the parameter $\Lambda$ is fixed by the scale of the chiral
symmetry spontaneous breaking, $\Lambda\simeq$ 0.7 -- 1.5 GeV,

\noindent
ii) the quark core radius of the nucleon should be in the region
$b_N\simeq$ 0.45 -- 0.6 fm,

\noindent
iii) the constituent quark mass should be
$m_q\simeq$ 200 -- 400 MeV.

In table \ref{table:paraset}, we show possible sets of parameters. The sets
I--III correspond to the inclusion of all types of the q--q forces, while
in set IV, there are no chiral fields inside the baryons, i.e.\
no $\pi$-- and $\sigma$--exchange between quarks.
This selection does not
give a systematic study of how the results depend on the
chosen parameters, but is meant to show some
qualitative features of our results.

Increasing $b_N$ up to $b_N$ = 0.6 fm in the above
discussed parameter fitting scheme, we get a too small constituent
quark mass, $m_q\simeq {\rm 170 MeV}$.

We also show in this table the model mass
for the two--quantum positive parity
J$^P$=$\frac{1}{2}^+$,T=$\frac{1}{2}$ resonance. This mass is calculated by
diagonalisation of our Hamiltonian with already fixed parameters in the
space including two possible two--quantum excitations
$\vert N_1\rangle$ and $\vert N_2\rangle$ with L=0 (see ref.\cite{gno}
and appendix A).

We see that this mass is essentially higher than the mass of the
Roper resonance. The Roper resonance is a longstanding problem in
the description of baryons within the constituent quark model.
We would like to remark at this point, that there are some new
ideas \cite{rig} for the  CQM, that seem to overcome
this problem.

Set V finally gives a set of parameters, determined from the following
fitting scheme, which will give the
dibaryon results closest to the "experimental value".
To obtain these parameters, we fix as input the constituent quark mass $m_q$,
the cut--off
$\Lambda$ and the nucleon hadronic size $b_N$ to determine $\alpha_s,a_c,C$
by the $\Delta$--N mass splitting, the nucleon
stability condition, and the nucleon mass respectively.
In the table we see
the small confinement strength $a_c\simeq$ 25 MeV/fm,
which goes together with a bigger quark core size of
the nucleons $b_N$=0.6 fm. These parameters emerge also naturally,
if the nucleon
quark core size $b_N$ is determined by the nucleon stability condition.
A similar parameter set is for example used to describe the
nucleon magnetic moments in \cite{buc}.
However, it is not possible to simultaneously describe the
one--quantum excitations with these parameters.

\bigskip
So, with fixed parameters, the effective 6--quark Hamiltonian is
given by
\dyy
H^{(6q)} = \sum_{i=1}^6\bigg( m_q+{{\bf p}_i^2\over 2 m_q}\bigg)
-{{\bf P}^2_{cm}\over 12 m_q}
\; +\; \sum_{i<j}^6 V^{Int}_{ij} \qquad ,\label{eq:ham}
\dzz
where
\dyy
V^{Int} = V^{Conf} + V^{OGE} + V^\pi + V^\sigma \qquad .
\label{eq:interham}\dzz
This Hamiltonian contains interactions of two particles only, so the
two--particle fractional parentage coefficients (fpc), which will be
reviewed below, are particularly well suited to evaluate
matrix elements for two--body interactions of the above type.

\section{The Translationally Invariant Shell Model basis (TISM)}

The harmonic--oscillator basis is highly successful in
describing the baryon spectrum.
Let us recall, why this is so. The reason is, of course, that the
quark--quark harmonic oscillator attraction simulates well the
quark--quark confinement forces in an 3--quark color--singlet object.
Indeed, the color--singlet 3--quark wave function
\dyy
[1^3]_C \equiv \;\sing
\label{eq:singlet}\dzz
is antisymmetric in color space in any quark pair. Since
\dyy
\langle \;\,\asym \vert \lilj \vert \;\,\asym \rangle
= -\frac{8}{3} \quad ,
\label{eq:antisym}\dzz
the confinement potential of eq.(\ref{eq:conf})
is attractive in any quark--quark pair.
Of course, historically, the confinement
potential was just constructed as to provide an attraction.

In the six--quark system, the situation is very different. Here the color part
is characterized by the mixed permutational symmetry
\dyy
[2^3]_C  \equiv \;\singl \; .
\label{eq:sixcolor}\dzz
It means, that there are both antisymmetrical and symmetrical pairs.
But in the symmetrical ones, instead of attraction, the potential
(\ref{eq:conf}) leads to repulsion, since
\dyy
\langle \;\,\sym \vert \lilj \vert \;\,\sym \rangle
= \frac{4}{3} \quad .
\label{eq:sym}\dzz
So, the confinement forces (\ref{eq:conf})
try to build the six--quark system in
such  a way as to divide it into two color--singlet 3--quark clusters.

It is not possible to simulate this behaviour with only one harmonic oscillator
six--quark configuration.
In the harmonic oscillator wave function we have an attraction in each
quark--quark pair\footnote{Due to the same reason,
the bag model for a six--quark
system is conceptually not correct.}.
However, one can use the 6q harmonic oscillator as a basis to diagonalize
the Hamiltonian eqs.(\ref{eq:ham}),(\ref{eq:interham}). In this case, it is
intuitively clear, that the confinement force will
mix the different 6q--harmonic oscillator states in such a way as to
provide in the final six--quark wave function a color--singlet --
color--singlet clusterization.
If the baryon--baryon asymptotics with well separated baryons (e.g.\ for the
deu\-teron) exists in some six--quark system, the harmonic oscillator basis
is not efficient to describe this system at all range, since we would need
a tremendous number of highly excited harmonic oscillator
configurations in this case\footnote{But
if the six--quark harmonic oscillator states at small distances are combined
with the corresponding clusterized configuration like
$\hat{{\cal A}} \{ B_1(1,2,3)B_2(4,5,6)\,\chi(\vec r)\}$
at medium and large range in one variational task, we do not need much
harmonic oscillator states at small distances. Such a basis is essentially
more flexible than the one--channel RGM basis. This variational program was
realized in part in the investigation of the $NN$ problem
in refs.\cite{ok,zbf}.}.

In the 6q--system with quantum numbers J$^P$=0$^-$, T=0, the nucleon--nucleon
clusterization is suppressed due to the Pauli principle on the nucleon level.
On the other hand, if there is a deeply bound state with respect to the
N(939)+N$^\ast$(1535) threshold, the
harmonic oscillator basis could be quite successful in describing such a
system.

The shell--model
basis allows, as we will see in more detail in the following, to respect
the Pauli principle, i.e.\ the antisymmetrization of the total wavefunction,
at any level of the calculation, so that effects arising due to the
Pauli--principle are naturally included in the formalism. This is
very important for interquark distances around or smaller than
$1$ fm, which are expected for a dibaryon.
The importance of the Pauli principle is well known from the NN--system.
There, the short range repulsion at distances smaller $1$ fm is
essentially due to quark exchanges, caused by the Pauli principle on the
quark level \cite{oy}.

\medskip
Furthermore, another advantage of this Hamiltonian lies in the fact, that the
center--of--mass (cm) motion is removed properly, so as to exclude
spurious states from the harmonic oscillator basis.

The price we pay for
the exact inclusion of the Pauli principle and
unambigious
exclusion of spurious states in our basis, is of course the uncertainties
induced by the non--relativistic description and by the effective nature of
our Hamiltonian, which is not derived from first  principles,
but is assumed to simulate the dominant
features of low--energy QCD.

\medskip
We remind the reader that the harmonic oscillator six--particle states are
exact solutions of the
Translationally Invariant
Shell Model (TISM) Hamiltonian, ref.\cite{kss}.
\daa
H^{(A=6)} &=& \sum_{i=1}^A {{\bf p}_i^2\over 2 m_q}
-{{\bf P}_{cm}^2\over 2A m_q} +{1\over 2}m_q\omega^2
\sum_{i=1}^A \big( {\bf r_i -R_{cm}}\big)^2
\nonumber\\
&=& \sum_{i=1}^A {{\bf p}_i^2\over 2 m_q}
-{{\bf P}^2_{cm}\over 2A m_q} +{m_q\omega^2\over 2A}\;
\sum_{i<j=1}^A \big( {\bf r_i -r_j}\big)^2 \quad .
\label{eq:smh} \dbb

\smallskip
The harmonic oscillator--eigenstates of this Hamiltonian
\dyy
\vert {\rm h.o.-state}\rangle =
\vert A=6\, ,N\, [f]_X \, (\lambda\mu ), L,S,T\,
,\alpha \rangle
\label{eq:smbasis}\dzz
are classified (cf.\ ref.\cite{kss}) by the following set of quantum numbers;
$N$ indicates the number of internal excitation quanta, the
Young--pattern $[f]_X$ determines the spatial permutational symmetry, the
Elliot symbol $(\lambda\mu )$ gives the $SU(3)$ harmonic oscillator multiplet,
$L,S,T$ are the total orbital angular momentum, total spin and
total isospin. The total spin $S$ is uniquely
connected with the Young pattern $[f]_S$ for the $SU(2)_S$ representation
by $S= (f_1 -f_2)/2$ where $f_1$ and $f_2$ are the first and second rows
in the Young pattern $[f]_S$. The same applies to the total isospin $T$.
The unambigious definition of a h.o.--state requires, as will be shown
in a moment, additional quantum numbers,
here collectively denoted by $\alpha$.

\medskip
In conventional nuclear physics (dealing with  nucleons),
due to the Pauli principle, the spatial Young pattern $[f]_X$ must be
conjugate to the $SU(4)_{ST}$ Young pattern $[f]_{ST}$ in order
to provide the total antisymmetry (i.e.
$[f]_X = [\tilde f]_{ST}$, where the tilde implies the Young
pattern reversed with respect to the main diagonal).

In the quark model (with two flavours), due to the additional degree of freedom
{\sl color}, the classification is more complicated. For a six--quark
system, the full permutational color symmetry is described by the $[2^3]_C$
color Young pattern ($SU(3)_C$ color singlet).
To identify unambigiously the six--quark state we have to use additional
quantum numbers, because the inner product of the $SU(3)_C$ color--singlet
representation and the $SU(2)_S$  spin representation (or the $SU(2)_T$)
contains in general more than one representation of the $SU(6)_{CS}$ group
\cite{jaf,ost,sk}, $SU(6)_{CS}\supset SU(3)_C\times SU(2)_S$
(or $SU(6)_{CT}$).
The spatial symmetry $[f]_X$ determines uniquely only the $SU(12)_{CST}$
permutational symmetry, $[f]_X = [\tilde{f}]_{CST}$; so we need in addition
$[f]_{CS}$ or $[f]_{CT}$.
Other reduction chains are also possible, for example with the intermediate
spin--isospin $SU(4)_{ST}$ symmetry \cite{ha},
or with the intermediate spin--spatial
symmetry \cite{obu}, etc. All these reduction chains are equivalent, and the
choice is determined by  convenience (for  different
types of  quark--quark forces  different reduction chains are
convenient).

In our preliminary communication \cite{gbf}, we have used the $SU(6)_{CS}$ and
$SU(4)_{ST}$ symmetries to classify all possible states. In the present paper
the choice of the intermediate spin--spatial permutational symmetry $[f]_{XS}$
is entirely motivated by the available tables of two--particle fractional
parentage coefficients in refs.\cite{obu}.

\vspace{.5cm}
\noindent
{\bf Classification of the states with J$^P$=0$^-$, T=0.}

\medskip
We get the negative parity only if our states contain an odd number of
harmonic oscillator excitations, i.e.\
the possible number of excitation quanta $N$ is restricted to
\dyy
N=1,3,5,...
\label{eq:excinumber}\dzz

There is only one state with N=1, compatible
with J$^P$=0$^-$, T=0:
\dyy
\vert N=1, [51]_X, (\lambda\mu)=(10), L=1, S=1, T=0, [42]_{XS}\rangle\;\; .
\label{eq:groundstate}\dzz
In this case, all possible intermediate
permutational symmetries (within different reduction chains)
are determined simultaneously and uniquely
\dyy
[321]_{CS}\; ,\; [321]_{ST}\; ,\; [2211]_{CT}\; ,\; [321]_{XT}\;
,\; [42]_{XS}\; ,\; [321]_{XC}\; .
\label{eq:intermediate}\dzz
This harmonic--oscillator state (i.e.\ TISM--state) is uniquely connected with
the harmonic--oscillator shell model state $s^5p$ as follows:
\dyy
\vert s^5p\; [51]_X\; (10) L=1\rangle = \Phi_{000}(\vec {\bf R}_{cm})\cdot
\vert N=1\; [51]_X\; (10) L=1\rangle
\label{eq:nonspurious}\dzz
The shell model state $s^5p$ with the spatial symmetry $[6]_X$ is
a spurious one, since
\dyy
\vert s^5p\; [6]_X\; (10) L=1\rangle = \Phi_{111}(\vec {\bf R}_{cm})\cdot
\vert N=0\; [6]_X\; (00) L=0\rangle
\label{eq:spurious}\dzz

\bigskip
There are a lot of different states with three internal
excitation quanta.

The full classification of the orbital parts
with N=3 and the spatial symmetries $[42]_X, [411]_X, [33]_X, [321]_X,$
and $[3111]_X$
was done in Ref.\cite{kss}. In addition we have also the states with
$[51]_X$ spatial symmetries (both $[6]_X$ and $[51]_X$ are also
allowed in the quark--model, but the state with N=3, $[6]_X$ is a spurious
one).

However, not all of these orbital
states are compatible with the color, spin and isospin quantum
numbers of the $d'$. In table \ref{table:states} we present all  possible
configurations
with N=3, compatible with the quantum numbers
J$^P$=0$^-$, T=0.

As mentioned above, all different reduction chains are
equivalent. For example, for the two reduction chains
\daa
SU(12)_{CST} &\supset & SU(6)_{CS}\times SU(2)_T
\supset SU(3)_{C}\times SU(2)_S\times SU(2)_T \quad ,
\label{eq:reductioncs}\\
SU(12)_{CST} &\supset & SU(3)_{C}\times SU(4)_{ST}
\supset SU(3)_{C}\times SU(2)_{S}\times SU(2)_T\quad ,
\label{eq:reductionst}\dbb
the intermediate $SU(6)_{CS}$ or $SU(4)_{ST}$ symmetries are needed
respectively to define the shell--model states uniquely. They are given
in line 4 and 5 of table \ref{table:states}
(see for more details ref.\cite{gbf}).

The reduction chain we finally adopt in our calculation, decouples color and
isospin from the remaining spin--orbital symmetry.
\daa
[1^6]_{XCST} &\supset & [33]_T\times\, [f'']_{XCS} \supset [33]_T\times
[222]_C\times\, [f]_{XS} \nonumber\\
&\supset & [33]_T\times [222]_C\times [f]_S\times\, [f']_X
\label{eq:reductionxs}\dbb
In the sixth line of table \ref{table:states},
we give the intermediate $[f]_{XS}$ symmetry,
needed in this classification scheme.

The rules for inner products of Young patterns for certain permutational
symmetries can lead to the situation, that a
certain product symmetry occurs more than once.
In these cases we need an additional
quantum number (in table \ref{table:states},
these quantum numbers are indicated  by
subscripts) to distinguish these different states.

The last line of table \ref{table:states}
gives the number of states in each column, all in all,
we find 31 TISM--states with N=3 and the quantum numbers of the d'.
Even if it were possible to include all these 31 excited states in our
calculation (this would give a $32\times 32$ matrix problem),
this would not be a complete basis, since we are not taking higher
excitations with $N= 5,7,...$ into account.
For a qualitative estimate of the effect of
configuration mixing with the N=3 states,
we choose the 10 states in the third row of table \ref{table:states},
characterized by
\dyy
N=3, \, [42]_X,\, L=1, S=1, T=0 \quad ,
\label{eq:excited}\dzz
in order to build a basis of 11 states for our calculation.
The notation for any of these ten states is analogous to
eq.(\ref{eq:groundstate}) and can be easily written down
with the help of table \ref{table:states}.

\smallskip
The reason why we choose these states is, that there are the two
most "symmetrical" (i.e.\ the longest) Young patterns $[f]_{CS}$ and
$[f]_{ST}$ among these states. We remind the reader, that the more
symmetrical pairs in color--spin--space we have,
the more attraction in the six--quark
system we get, due to the color--magnetic
$\lilj\sisj$ forces in eq.(\ref{eq:ogep}).

This is well seen from the matrix elements
\dyy
\langle [f_{ij}]_C \times [f_{ij}]_S \vert\;\lilj\sisj\;
\vert [f_{ij}]_C \times [f_{ij}]_S \rangle = \left\{ \begin{array}{llll}
\frac{4}{3}, & {\rm if} \; [2]_C, & [2]_S, & [2]_{CS} \\
8, & {\rm if} \; [11]_C, & [11]_S, & [2]_{CS} \\
-4, & {\rm if} \; [2]_C, & [11]_S, & [11]_{CS} \\
\frac{-8}{3}, & {\rm if} \; [11]_C, & [2]_S, & [11]_{CS}
\end{array} \!\!\right.
\label{eq:colorspin}\dzz
The same situation takes place for the pion--exchange forces (\ref{eq:opep}).
Here, one has
\dyy
\langle [f_{ij}]_S \times [f_{ij}]_T \vert\;\titj\sisj\;
\vert [f_{ij}]_S \times [f_{ij}]_T \rangle = \left\{ \begin{array}{llll}
1, & {\rm if} \; [2]_S, & [2]_T, & [2]_{ST} \\
9, & {\rm if} \; [11]_S, & [11]_T, & [2]_{ST} \\
-3, & {\rm if}\; [11]_S, & [2]_T, & [11]_{ST} \\
-3, & {\rm if}\; [2]_S, & [11]_T, & [11]_{ST}
\end{array} \!\!\right.
\label{eq:spinisospin}\dzz
So, at quark--quark distances larger than, for example, $1$ fm, where the first
term in eq.(\ref{eq:opep}) dominates over the third one, one has an attraction
due to central $\pi$--exchange forces in antisymmetrical $ST$--pairs.
At small quark--quark distances (less than $1$ fm), the cut--off in
eq.(\ref{eq:opep}) dominates,
and one has an attraction in symmetrical $ST$--pairs.
Since the characteristic size of our system is less than $1$ fm, one has
an attraction in symmetrical $ST$--pairs.

\vspace{.5cm}
\noindent
{\bf The fpc (fractional parentage coefficients) technique}

\medskip
Here, we cite briefly the definition and the application of the
fractional parentage expansion (fpc), refering the reader
to the corresponding literature \cite{kss}--\cite{obu} for more details.

Each six--quark harmonic oscillator configuration can be presented by
means of the fractional parentage technique and the
Talmi--Moshinsky transformation as a superposition  of various
four--quark $\times$ two--quark $\times$ relative motion  components:
\dyy
\vert N \;\alpha \; LSJT\rangle =
\sum_{B_1,B_2,n,l} \Gamma^\alpha_{B_1,B_2,n,l} \bigg\{
\phi_{B_1}(1,2,3,4)\,\phi_{B_2}(5,6)\,\phi_{nl}(\vec r)\, :LSJT\bigg\}
\label{eq:cfpdef}\dzz
Here, $\alpha$ stands for the necessary set of additional
quantum numbers,  including
$[f]_X ,(\lambda\mu ), [f]_{XS},$ etc.\ to define unambigiously the
six--quark state, and $\Gamma^\alpha_{B_1,B_2,n,l}$ is the so--called
fractional parentage coefficient (fpc) in the TISM.

$\phi_{B_1}$ and $\phi_{B_2}$ are the TISM states for the first four
particles and for the last two ones respectively. $\phi_{nl}(\vec r)$
is the n--quantum harmonic oscillator function with
$$\vec r =\frac{\vec r_1+\vec r_2+\vec r_3+\vec r_4}{4} -
\frac{\vec r_5+\vec r_6}{2}\quad .$$
The summation in eq.(\ref{eq:cfpdef})
is carried out over all possible internal states of clusters $B_1$ and $B_2$
and their relative motion $n,l$, provided that
$$N_1 + N_2 + n = N \qquad .$$
The specific feature of expansion (\ref{eq:cfpdef}) is, that the
antisymmetric six--quark wave function is expanded into the sum of orthogonal,
but not fully antisymmetric terms. Each term is antisymmetric only within
the clusters $B_1$ and $B_2$.

\medskip
By use of expansion (\ref{eq:cfpdef}), the interaction energy matrix elements
of a pure two--body potential
$H_{int} = \sum_{i<j}^N V_{ij}$
can directly be evaluated:
\daa
\langle\!\! &N& \!\alpha
\; LSJT\vert\; H_{int}\;\vert N\;\alpha '\; L'S'JT\rangle
= \frac{6(6-1)}{2} \times \label{eq:cfpmatrix} \\
\! &\times& \!\!\!\!\sum_{B_1,B_2,\tilde B_2,n,l}
\!\Gamma^\alpha_{B_1,B_2,n,l} \,
\Gamma^{\alpha '}_{B_1,\tilde B_2,n,l} \times\!\!
\left( \begin{array}{l}
{\rm angular} \\ {\rm momentum} \\ {\rm recoupling} \\ {\rm matrix}
\end{array} \right) \!\!\times
\langle \phi_{B_2}(5,6)\vert V_{56}\vert \phi_{\tilde B_2}(5,6) \rangle
\nonumber\dbb
The expression (\ref{eq:cfpmatrix}) is written in symbolical form to
avoid bulky $6j$-- and $9j$--symbols appearing due to the necessary recoupling
of the angular momenta when calculating a contribution of the non--central
forces. Here, a summation over all intermediate momenta is also assumed. For
central forces, the recoupling matrix is absent.

\medskip
It will not be our task here to rederive the rules of how
to construct the two--particle fpc's for the 6--particle system.
General considerations for the construction of fpc's can be found in various
publications, as for example in \cite{kss}--\cite{obu} for the 6--quark
system, and the fpc's (or more precisely the scalar factors)
needed in our case are tabulated in \cite{ost,sk} and
\cite{obu}.
It is shown in these articles (and in the articles cited therein), that
for a given reduction chain, the total fpc--coefficient factors out in a
product of several scalar factors of the Clebsch--Gordan coefficients
of the corresponding unitary groups, each one associated with a
single step of the reduction.

In our case, the corresponding fpc are given as a product of scalar factors
\dyy
\Gamma^\alpha_{B_1,B_2,n,l}
=SF^U_{XSC\cdot T}\cdot SF^U_{XS\cdot C}\cdot SF^U_{X\cdot S}\cdot SF^U_C
\cdot\Gamma_{TISM}\quad .
\label{eq:scalarfactor}\dzz

The first factor in eq.(\ref{eq:scalarfactor}) is the weight factor
and is determined only by the dimensions of the irreducible representations
of the permutational group in isospin space for six particles (Young pattern
$[f]_T$) and for the first four particles in eq.(\ref{eq:cfpdef}) (Young
pattern $[f_1]_T$):
\dyy
SF^U_{XSC\cdot T} = \sqrt{\frac{{\rm dim}[f_1]_T}{{\rm dim}[f]_T}} \; .
\label{eq:weightfactor}\dzz

The next two factors in eq.(\ref{eq:scalarfactor}) can be extracted from the
tables 2 -- 5 in ref.\cite{obu}.
The color scalar factor (i.e.\ the scalar part of the $SU(3)_C$ Clebsch--Gordan
coefficient in the reduction $SU(3)_C\supset O(3)_C$) can be found in table 1
of ref.\cite{ost}.

Finally, the last factor in eq.(\ref{eq:scalarfactor}) $\Gamma_{TISM}$,
i.e.\ the orbital part of the fpc, can be found in table 2k of the
first paper from \cite{kss}.
This table contains all the necessary coefficients for
N=3, $[f]_X =[42]_X$ six--particle harmonic oscillator configurations.
For the lowest configuration, i.e.\ for
$\vert N=1\, [51]_X\, (\lambda\mu )=(10)\, L=1,S=1,T=0\rangle$, these
coefficients are trivially calculated, and one finds:
\daa
\langle N\! =\! 1 [51]_X\, (10)\, L\! =\! 1 \vert N_1=0 [4]_X(00)\, L_1=0\, ;
nl=00\, , N_2=1 [11]_X\, L_2=1 \rangle\! &=&\! -1 \nonumber\\
\langle N\! =\! 1 [51]_X\, (10)\, L\! =\! 1 \vert N_1=0 [4]_X(00)\, L_1=0\, ;
nl=11\, , N_2=0 [2]_X\, L_2=0 \rangle\! &=&\! -1 \nonumber\\
\langle N\! =\! 1 [51]_X\, (10)\, L\! =\! 1 \vert N_1=1 [31]_X(10)\, L_1=1\, ;
nl=00\, , N_2=0 [2]_X\, L_2=0 \rangle\! &=&\! 1 \nonumber\\
\label{eq:groundorbit}\dbb

\section{Results for the single N=1 configuration}

As our full six--quark Hamiltonian of eq.(\ref{eq:ham})
for the dibaryon comprises not only
confinement, but also effective gluon--, pion-- and sigma--exchange,
it is very different from the harmonic
oscillator Hamiltonian of eq.(\ref{eq:smh}).

In our previous paper \cite{gbf}, we treated the difference between the
effective Hamiltonian of eq.(\ref{eq:ham}) and the harmonic--oscillator
Hamiltonian of eq.(\ref{eq:smh}) as a residual interaction.
The corresponding diagonal matrix element for the lowest possible
configuration, given by eq.(\ref{eq:groundstate})
\dyy
M^{(N=1)} \! =\! \langle N=1, [51]_X (10) {\rm LSTJ=1100} \vert H^{(6q)}
\vert N=1, [51]_X (10) {\rm LSTJ=1100} \rangle \; ,
\label{eq:groundresult}\dzz
can be found in ref.\cite{gbf} and also in appendix B.

As we already mentioned in the previous sections, the confinement forces
of eq.(\ref{eq:conf}) work towards the color--singlet + color--singlet
clusterization in the 6q--system. In our case, that could be
NN$^\ast_{\frac{1}{2}^-}$(1535) and
NN$^\ast_{\frac{3}{2}^-}$(1520) clusterizations.
We can exclude higher excited nucleon states or N$^\ast$N$^\ast$
clusterizations, since their threshold energies would be quite high.
However, the
NN$^\ast_{\frac{1}{2}^-}$(1535) is highly preferable compared to the
NN$^\ast_{\frac{3}{2}^-}$(1520). The reason is, that the latter cluster
component in the J$^P$=0$^-$, T=0 six--quark system is not
compatible with the lowest possible
harmonic oscillator configuration of eq.(\ref{eq:groundstate}). Indeed, in the
NN$^\ast_{\frac{3}{2}^-}$(1520) system, the relative motion angular momentum
must be $l=2$ to provide total angular momentum and parity J$^P$=0$^-$
of the six--quark state.
Such an angular momentum $l$ needs a non--zero number of harmonic oscillator
quanta, $n=2,4,\ldots$, corresponding to the relative motion.
So, together with the one quantum from the internal N$^\ast$ excitation,
the total number of harmonic oscillator quanta would be at least 3.
The NN$^\ast_{\frac{1}{2}^+}$(1440) component is also incompatible with the
N=1 six--quark state.
Thus, the "dynamical" threshold for a possible J$^P$=0$^-$, T=0 dibaryon
is N(939)+N$^\ast$(1535).

\smallskip
It is instructive to compare the excess energy over threshold in our case,
i.e.\ $\delta M= M^{(N=1)}-(m_N+m_{N^\ast})$,
with the corresponding value in the deuteron--like six--quark system
J$^P$=1$^+$,T=0 ($\, ^3S_1\; NN$--wave),
$\delta M_{s^6}= M_{s^6}-2m_N$. In the latter case, as we know,
there exists no compact 6q--state and the bound system (deuteron)
consists of two weakly bound nucleons, which with overwhelming
probability are far from each other.

\smallskip
We see from table \ref{table:excess}
that $\delta M$ is very similar to
$\delta M_{s^6}$.
This maybe a hint,
that there could be a weakly bound state (below the N(939)+N$^\ast$(1535)
threshold) in the
J$^P$=0$^-$,T=0 system.
The reason is quite clear.
There are more symmetrical pairs with CS--symmetry, $[321]_{CS}$, in
the J$^P$=0$^-$,T=0 system, than with the $[2^3]_{CS}$ symmetry corresponding
to the deuteron--like $s^6$ configuration.
But the $ST$ Young pattern for the
J$^P$=1$^+$,T=0,
$s^6$ configuration, $[33]_{ST}$,
is more symmetrical than $[321]_{ST}$, inherent in the state
(\ref{eq:groundstate}) according to eq.(\ref{eq:intermediate}).
As a consequence the gain in energy coming
from the color-magnetic forces (prefering a symmetric CS--configuration)
is approximately compensated by the loss
in energy arising from the chiral-exchange interaction (which prefers
symmetrical ST--configurations). As a consequence, both systems
have very similar excess energies.

\smallskip
In table \ref{table:excess}, we have shown the excess $\delta M$ calculated
with the same harmonic oscillator parameter $b_6$ in the six--quark system
as for the baryons $b_6=b_N$. However, with fixed parameters of the
quark--quark interactions of eq.(\ref{eq:interham}), one should minimize the
"mass" eq.(\ref{eq:groundresult}) with respect to the harmonic oscillator
parameter $b_6$ in the six--quark trial wave function. The corresponding
value $b_6$ and the mass of the 6q--state $M^{T=0}_{N=1}$,
are shown in the 3$^{\rm rd}$ and 4$^{\rm th}$
column of table \ref{table:excess}.
A more detailed discussion of these results for different
parameter sets can be found in \cite{gbf}.

\section{Configuration mixing results}

Here, we report on the diagonalisation of the Hamiltonian (\ref{eq:ham}) in
the space including 11 harmonic oscillator configurations (the only state
 in the N=1 shell, and 10 states with a total number
of three harmonic oscillator quanta excited:
N=3,$[42]_X$, L,S,T=1,1,0). The quantum numbers and our labelling of
these states, as used in the calculation, are given in table
\ref{table:basisstates}.
All the needed matrix elements
\dyy
H_{ij} \; =\; \langle\, i\,\vert\; H\;\vert\, j\,\rangle
\qquad ;\qquad i,j=1\ldots 11
\label{eq:matrixelements}\dzz
can be found in appendix B.

\smallskip
It is important to mention the well known fact that a $\delta$--type
attractive two--body potential would result in a collapse in the three--body
system, if a complete basis were used.
Thus, the $\delta$--type forces in
eq.(\ref{eq:ogep}) have to be considered as an effective interaction, only
valid for a given finite basis.
We included the contact forces in perturbation theory and in the full
diagonalisation, and obtained within 2 to 3 MeV the same masses for the
calculated resonance J$^P$=0$^-$,T=0.

\smallskip
Since our basis is not complete, we improve its flexibility by using the
harmonic oscillator parameter $b_6$ in the six--quark basis functions
as a nonlinear variational parameter. More clearly, with fixed
parameters of the quark--quark interactions, we minimize the lowest eigenvalue
of (\ref{eq:matrixelements})
with respect to $b_6$. This means in particular,
that the harmonic oscillator parameters (and thus the root mean square radius)
of the baryons $b_N$ and in the dibaryon $b_6$ need not to be the same.

\medskip
Let us now discuss different contributions to the mass
from the Hamiltonian (\ref{eq:ham}).
Since the diagonalisation of eq.(\ref{eq:matrixelements}) is a very nonlinear
procedure, it is difficult to estimate the importances of the separate parts
of the Hamiltonian (\ref{eq:ham}). Nevertheless, diagonalising for
example the confinement matrix (table \ref{table:conf}) alone, one recognizes
that the mixing due to the confinement is essential. This result is
independent of the chosen parameter set, since the parameter dependence
factors out for the diagonalisation as is seen from eq.(\ref{eq:confi}) in
appendix B.

The same argument holds for the individual diagonalisation of the
color--Coulomb interaction eq.(\ref{eq:colcoul}).
In this context, it is important to see, that large mixing amplitudes
for the diagonalised system do not necessarily lead to a large gain in
energy by the mixing calculation. For the confinement for example, the
energy gain is rather important (around 100--150 MeV) whereas in the case
of the color--Coulomb interaction, we gain only about 30 MeV.

\smallskip
One should mention here, that the effect of all forces, that involve only
spatial and color degrees of freedom in their potentials, could be
studied more easily in the basis with intermediate $ST$--symmetry. Here, these
forces are diagonal with respect to the spin-isospin symmetry,
and we have no mixing between
states with different $ST$--symmetries for this part of the Hamiltonian.
This means that only two of the 10 chosen states from the N=3 shell with
$[321]_{ST},(\lambda\mu )=(11)$ and
$[321]_{ST},(\lambda\mu )=(30)$ would be mixed in a ST--basis
to the $N=1,[321]_{ST}$
state by the confinement, color--Coulomb and $\sigma$--exchange forces.
Using the ST reduction chain (\ref{eq:reductionst}), only the color--magnetic,
tensor and/or pion interactions would mix among states with different $ST$.

\smallskip
With all parts of the Hamiltonian (\ref{eq:ham}),
this analysis of the interplay of the different potential parts cannot be
done anymore. Of course, the difference in kinetic energy
from the N=1 and the N=3 shell of $\frac{1}{m_q b^2}\simeq$300 -- 500 MeV
is the most obvious reason for the rather moderate gain in energy as seen
from table \ref{table:excess}.
table \ref{table:excess} shows in its last 2 columns first the
harmonic oscillator length $b_6$, for which
the minimum of the mass is reached, and second the corresponding mass
eigenvalue
$M$ for the chosen parameter set.
This value should be compared (for the parameter sets I to IV) to the
N(939)+N$^\ast$(1535) threshold of $2474$ MeV.

Comparing the results with and without configuration mixing, we see that
the inclusion of the excited states shifts the mass down by about
$100$ -- $150$ MeV. In the best case (for set II), the mixing result
is $90$ MeV above the N(939)N$^\ast$(1535) threshold.
The inclusion of higher configurations makes the 6q--system larger,
and the confinement strength determines essentially the
size of the dibaryon $b_6$. Therefore, since parameter set V allows for
a large dibaryon of $b_6 = 1.25$ fm, and since the kinetic energy
is proportional $1/b^2$, we see qualitatively, why this set
gives the lowest "mass" of the dibaryon in our basis.
One can see in table \ref{table:paraset} that this
parameter set does not allow for a
description of the excited baryon N$^\ast$(1535).
Thus, we conclude that it is difficult
to describe the baryon spectrum and the dibaryon mass with the same
confinement parameters.

\section{The J$^P$=0$^-$, T=2 six--quark state as a candidate for the d'}

In this section, we discuss shortly the question of a
J$^P$=0$^-$, T=2 six--quark state, since from the very beginning, it is not
clear which state, T=0 or T=2, is lower in mass \cite{gar}.
Both possible states
can not decay into NN--channels due to the Pauli--principle.

\medskip
In our approach, following the classification (\ref{eq:reductionxs})
in section 3, we get
two states with one excitation quantum N=1 and the quantum
numbers J$^P$=0$^-$, T=2.
\daa
\vert 1\rangle &=& \vert N\! =\! 1,[51]_X,(10),
L\! =\! 1,S\! =\! 1,T\! =\! 2, [321]_{XS} \rangle \label{eq:state1} \\
\vert 2\rangle &=& \vert N\! =\! 1,[51]_X,(10),
L\! =\! 1,S\! =\! 1,T\! =\! 2, [42]_{XS} \rangle \label{eq:state2}
\dbb
With the other possible reduction chains (\ref{eq:reductioncs}) and
(\ref{eq:reductionst}), we get for (\ref{eq:reductionst}) the two states
$[42]_{ST}$ and $[321]_{ST}$, and for
(\ref{eq:reductioncs})
$[31^3]_{CS}$ and $[21^4]_{CS}$.
The necessary diagonal matrix--elements for the states (\ref{eq:state1}) and
(\ref{eq:state2}) are given in appendix C.

In columns 5 and 6 of
table \ref{table:excess}, we give the minimized expectation
values (\ref{eq:mass1}) and (\ref{eq:mass2}) for the lowest possible
states $\vert 1\rangle$ and $\vert 2\rangle$ in comparison with the result
for the N=1, J$^P$=0$^-$, T=0 state.
 From table \ref{table:excess} we see, that within the constituent quark model
both the J$^P$=0$^-$,T=0 and the
J$^P$=0$^-$,T=2 state are nearly degenerate.

The reason for this degeneracy
is clearly the competition between the color--magnetic part of
the gluon--exchange and the central pion--exchange contribution.
In analogy to the discussion in section 3, we see here, that the
T=0 state with its $[321]_{CS}$ symmetry provides more attraction from the
color--magnetic interaction, than the less symmetric $CS$--symmetries of the
T=2 states. The competing process is the pion--exchange part, which provides
for the $[42]_{ST}$ symmetry of the T=2 state more gain in energy, than in the
N=1, T=0, $[321]_{ST}$ state. So, at the end, the results depend very much on
the given parameter set, or more precisely on the relative
importance of the one--gluon-- and the pion--exchange.
Both six--quark systems
are quite close in energy, and from the point of view of the constituent
quark model neither the T=0 nor the T=2 state are energetically favoured.

\section{Conclusion}

We have investigated the six--quark system with the quantum numbers
J$^P$=0$^-$, T=0 within the constituent quark model with and without chiral
interactions ($\pi$-- and $\sigma$--exchange)
between quarks. The aim of this study was to look for a
possible dibaryon with these quantum numbers. It was argued \cite{bcs}, that
such a dibaryon is seen as a narrow peak in double charge exchange reactions
$(\pi^+,\pi^-)$ on various nuclei.

We have classified all possible six--quark states
J$^P$=0$^-$, T=0 with N=1 and N=3 harmonic oscillator excitations using
different reduction chains. We then have diagonalized
the microscopic Hamiltonian
including the linear quark--quark confinement, chiral interactions
between quarks, and the effective one--gluon exchange--potential in the
basis consisting of the lowest (N=1) state and the 10 presumably most
important configurations from the N=3 shell.
We have argued, that the calculated mass--eigenvalue should be
compared to the 2--baryon N(939)+N$^\ast$(1535) threshold.

We have found, that if one fits all parameters
to describe the nucleon and
its lowest resonances, the mass of such a "dibaryon" lies above the
N(939)+N$^\ast$(1535) threshold, when taking our restricted basis of 11 states.
Since we gain around 100 -- 150 MeV
by admixing the 10 chosen states, it seems not to be
unrealistic, that a more complete basis would lead to a "dibaryon mass"
below the N(939)N$^\ast$(1535) threshold of 2474 MeV.
So, our calculations do not deny the existence of a dibaryon with
J$^P$=0$^-$, T=0. Its mass could be some 10 MeV's below the NN$^\ast$
threshold (within the framework of our model).
But the discrepancy with the "experimentally found" mass is still large.

We cannot claim for sure, that the constituent quark model "excludes"
the interpretation of the peak in $(\pi^+,\pi^-)$ reactions as a signal
from the dibaryon with the above mentionned quantum numbers, since our
basis is not complete, and as a consequence, we lose probably some of the
possible short--range and few--quark correlations in the
six--quark system.
Another point to mention is the following:
The 6q--system differs
qualitatively from the description of a baryon in the sense, that the
6q--color symmetry is a mixed symmetry, in contrast to the fully antisymmetric
color--wavefunction of a baryon.
Probably, the confinement for a 6q--system (in addition to nuclear
medium effects in the nucleus)
differs qualitatively from the confinement mechanism
in the baryons.
At this point, we should add, that the
choice of a quadratic
or a so--called color--screened exponential \cite{zha}
rather than a linear confinement potential does not
change qualitatively our results.
If the assumption of a dibaryon resonance at 2065 MeV is confirmed,
we could probably improve our understanding of confinement in
multi--quark systems.

\smallskip
We have also analyzed a possible
J$^P$=0$^-$, T=2 dibaryon. The corresponding lowest shell model state lies
below the J$^P$=0$^-$, T=0 one, provided that the one--pion--exchange forces
play an essential role in the 6q--system.

Finally, we have shown, that we could describe an "observed dibaryon"
with a mass close to
$M_{d'}$= 2065 MeV, by reducing the confinement strength
$a_c$ (string tension), down to $a_c\simeq$ 25 MeV/fm \cite{buc}
(see parameter set V).
But, such a
confinement strength does not give the correct energy for the negative parity
one--quantum excitation
N$^\ast$(1535).
Thus, we observe two opposing tendencies within our model:
We need a small confinement strength (and therefore a large size
parameter $b_N$ for the nucleon) to
get a rather large dibaryon, which is consequently rather
light in mass. On the
other hand, we need a rather small nucleon size $b_N$ to describe the baryon
spectrum, especially the N$^\ast$ excited states.
This result suggests,
that this six--quark system is strongly influenced
by the confinement mechanism.
The above discrepancy can not be eliminated by another
choice of radial dependence of the
confinement potential (quadratic or "color--screened" exponential) alone.
If the d' dibaryon is confirmed, we suggest that
the color--structure $\lilj$ of the confinement potential should be modified
in multi--quark systems like the d'.
We should stress here once more, that it
seems impossible to describe  the nucleon resonances (the 3q--sector)
and this dibaryon resonance within the same model with the  same
parameters.

\bigskip
{\bf Acknowledgements:}
L.Ya.G.\ thanks the A.\ von Humboldt
Foundation for a fellowship and the Institute of Theoretical
Physics at the University of Tuebingen for their hospitality.
G.W.\ thanks the DFG Graduiertenkolleg "Struktur und Wechselwirkungen von
Hadronen und Kernen" for financial support (contract number Mu705/3).

\vspace{1cm}
\appendix{\large\bf Appendix A}

\setcounter{equation}{0}
\renewcommand{\theequation}{A.\arabic{equation}}
\setcounter{table}{0}
\renewcommand{\thetable}{A.\arabic{table}}

\medskip
In this appendix, we give the mass formulae for the baryon systems
\dyy
{\rm N(939), }\;\Delta{\rm (1232), }\;
{\rm and}\; {\rm  N}^\ast{\rm (1535),}\;{\rm N}^{\ast\ast}{\rm (1440)}
\label{eq:baryons}\dzz
within the constituent quark model, as described in the
parameter fitting section.

The following harmonic oscillator basis states
\daa
\mid N_0 > &=& \mid N=0, (\lambda \mu)=(00), [3]_X, L=0, S=\frac {1}{2},
T=\frac {1}{2}\; [3]_{ST} \rangle \; ,\nonumber\\
\mid \Delta_0 > &=& \mid N=0, (\lambda \mu)=(00), [3]_X, L=0, S=\frac {3}{2},
T=\frac {3}{2}\; [3]_{ST} \rangle \; ,\nonumber\\
\mid N^{(-)}_1 > &=& \mid N=1, (\lambda \mu)=(10), [21]_X, L=1, S=\frac {1}{2},
T=\frac {1}{2}\; [21]_{ST} \rangle \; ,\nonumber\\
\mid N^{(-)}_2 > &=& \mid N=1, (\lambda \mu)=(10), [21]_X, L=1, S=\frac {3}{2},
T=\frac {1}{2}\; [21]_{ST} \rangle \; ,\nonumber\\
\mid N_1 > &=& \mid N=2, (\lambda \mu)=(20), [3]_X, L=0, S=\frac {1}{2},
T=\frac {1}{2}\; [3]_{ST} \rangle \; ,\nonumber\\
\mid N_2 > &=& \mid N=2, (\lambda \mu)=(20), [21]_X, L=0, S=\frac {1}{2},
T=\frac {1}{2}\; [21]_{ST} \rangle \; ,
\label{eq:nstar2}\dbb
are needed for the calculation of the appropriate matrix elements
which determine the masses of
the baryons N(939), $\Delta$(1232), N$^\ast$(1535) and
N$^{\ast\ast}$(1440) in our chosen parameter
fitting scheme. The ground states are denoted by the subscript "0".
The detailed wave functions can be found in the second paper of
\cite{gno}.

Our Ansatzes for the different baryons are:
\daa
\vert N \rangle &=& \vert N_0 \rangle \quad :\quad
\vert\Delta \rangle = \vert \Delta_0 \rangle \nonumber \\
\vert N^\ast \rangle &=& \lambda\;\vert N_1^{(-)} \rangle +
                         \mu\;\vert N_2^{(-)} \rangle \nonumber \\
\vert N^{\ast\ast} \rangle &=& \alpha \;\vert N_1 \rangle +
                         \beta\;\vert N_2 \rangle
\label{eq:ansaetze}\dbb

To be rather short, we give here all necessary (diagonal and
off--diagonal) matrix elements
in their analytical form in one table.
\daa
h^{N,\Delta ,N^{(-)},N^{\ast\ast}}_{ij} &=&
3m_q\,\delta_{ij} + k_{ij}\,\frac{1}{m_q b^2}+a_{ij}\,\frac{a_c b}{\sqrt{2\pi}}
-2C\,\delta_{ij} + b_{ij}\,\frac{\alpha_s}{\sqrt{2\pi}\, b} + \nonumber\\
&+& (ce_{ij}+cm_{ij})\frac{\alpha_s}{\sqrt{2\pi}m_q^2 b^3} +
V^\pi_C + V^\pi_T + V^\sigma_C
\label{eq:general}\dbb
The notations for
the pion central--contribution are the same as in (\ref{eq:pionic})
and for the $\sigma$--exchange the notations are the same as in
(\ref{eq:sigisigi}). As in appendix B, we do not give the lengthy
formulae for the pion--tensor contribution in explicit form.

\begin{center}
\begin{tabular}{c||c||c|c|ccc|c|c}
 & & k & a & b & ce & cm & p,p1,p3,p5 & s,s1,s3,s5 \\ \hline\hline
N & $h^N_{00}$  &
  $\frac{3}{2}$ & 8 & --4 & 1 & $\frac{-2}{3}$ & 5,1,0,0 & --3,1,0,0 \\
\hline
$\Delta$ & $h^\Delta_{00}$  &
  $\frac{3}{2}$ & 8 & --4 & 1 & $\frac{+2}{3}$ & 1,1,0,0 & --3,1,0,0 \\
\hline
 & $h^{N^{(-)}}_{11}$  &
  2 & $\frac{28}{3}$ & $\frac{-10}{3}$ &
  $\frac{1}{2}$ & $\frac{-1}{3}$ & $\frac{1}{2}$,5,--1,0 &
  $\frac{-3}{2},1,\frac{1}{3},0$ \\
N$^\ast$ & $h^{N^{(-)}}_{22}$  &
  2 & $\frac{28}{3}$ & $\frac{-10}{3}$ &
  $\frac{1}{2}$ & $\frac{1}{3}$ & $\frac{1}{2}$,1,--1,0 &
  $\frac{-3}{2},1,\frac{1}{3},0$ \\
 & $h^{N^{(-)}}_{12}$  &
  0 & $\frac{2\sqrt{2}}{3}$ & $\frac{\sqrt{2}}{3}$ &
  $\frac{-1}{2\sqrt{2}}$ & $\frac{-1}{3\sqrt{2}}$ &
  $\frac{-1}{2\sqrt{2}}$,1,1,0 &
  $\frac{3}{2\sqrt{2}},1,\frac{-1}{3},0$ \\
\hline
 & $h^N_{11}$  &
  $\frac{5}{2}$ & 10 & $\frac{-11}{3}$ & $\frac{5}{4}$ & $\frac{-5}{6}$ &
  $\frac{5}{2},\frac{5}{2},-1,\frac{1}{6}$ &
  $\frac{-3}{2},\frac{5}{2},-1,\frac{1}{6}$ \\
N$^{\ast\ast}$ & $h^N_{22}$  &
  $\frac{5}{2}$ & $\frac{31}{3}$
  & $\frac{-19}{6}$ & $\frac{5}{8}$ & $\frac{-5}{12}$ &
  $\frac{1}{4},\frac{25}{2},-7,\frac{5}{6}$ &
  $\frac{-3}{4},\frac{5}{2},\frac{-1}{3},\frac{1}{6}$ \\
 & $h^N_{12}$  &
  0 & 0 & 0 & 0 & $\frac{1}{3\sqrt{2}}$ & $-\sqrt{2},\frac{1}{2},-1,
  \frac{1}{6}$ & 0 \\
\end{tabular}
\end{center}

\vspace{0.5cm}
\appendix{\large\bf Appendix B}

\setcounter{equation}{0}
\renewcommand{\theequation}{B.\arabic{equation}}
\setcounter{table}{0}
\renewcommand{\thetable}{B.\arabic{table}}

\medskip
In this appendix, we present all matrix elements
\daa
H_{ji} \equiv H_{ij} &=& \langle\, i\,\vert\; H\;\vert\, j\,\rangle
\qquad ; \qquad i,j=1\ldots 11 \nonumber\\
&=& 6m_q\;\delta_{ij} + \big( E^{kin} +
V^{conf} + V^{OGEP} + V^\pi + V^\sigma \big)_{ij}
\label{eq:matrixelement}\dbb
for the different parts of the Hamiltonian eq.(\ref{eq:ham}).
in the tables \ref{table:conf} through \ref{table:sigma},
excluding the
pion tensor contribution again.
As mentioned above, the labelling of the basis states, as they are
used here, is given in table \ref{table:basisstates}.

\medskip
The kinetic energy contribution is given by
\dyy
E^{kin}_{ij} = \left\{ \begin{array}{ll}
\frac{17}{4}\,\frac{1}{m_q\, b^2} & {\rm if} \; i=j=1 \\
\frac{21}{4}\,\frac{1}{m_q\, b^2} & {\rm if} \; i=j=2\ldots 11 \\
0                                 & {\rm if} \; i\neq j
\end{array}\right.
\label{eq:kinetic}\dzz

The coefficient $a_{ij}$ of the confinement contribution
\dyy
V^{conf}_{ij} =
\langle i\vert \sum_{k<l}^6 - \frac{\lkll}{4}\, (a_c r -C)
\vert j\rangle
= a_{ij}\,\frac{a_c b}{\sqrt{2\pi }} -4C\,\delta_{ij}
\label{eq:confi}\dzz
is given by

\medskip
Tables \ref{table:coco} -- \ref{table:cot}
give the different parts of the one--gluon exchange potential,
starting with the color Coulomb contribution
\dyy
V^{CC}_{ij} =
\langle i\vert \sum_{k<l}^6 \frac{\alpha_s}{4}\,\lkll\,\frac{1}{r}
\vert j\rangle
= - b_{ij}\,\frac{\alpha_s}{\sqrt{2\pi }b}
\label{eq:colcoul}\dzz

\medskip
The next table gives the so--called color--electric part
\dyy
V^{CE}_{ij} =
\langle i\vert \sum_{k<l}^6 -\frac{\alpha_s}{4}\,\frac{\pi}{m_q^2}\,
\lkll\,\delta (\vec r)
\vert j\rangle
= ce_{ij}\,\frac{\alpha_s}{\sqrt{2\pi } m_q^2 b^3}
\label{eq:colel}\dzz

\medskip
This is the color--magnetic contribution
\dyy
V^{CM}_{ij} =
\langle i\vert \sum_{k<l}^6 -\frac{\alpha_s}{6}\,\frac{\pi}{m_q^2}\,
\lkll\sksl\,\delta (\vec r)
\vert j\rangle
= cm_{ij}\,\frac{\alpha_s}{\sqrt{2\pi } m_q^2 b^3}
\label{eq:colmag}\dzz

\medskip
The next table gives the gluon tensor contributions
\dyy
V^{GT}_{ij} =
\langle i\vert \sum_{k<l}^6 -\frac{\alpha_s}{16 m_q^2}\,
\lkll\,\hat S_{kl}\,\frac{1}{r^3}
\vert j\rangle
= ct_{ij}\,\frac{\alpha_s}{\sqrt{2\pi } m_q^2 b^3}
\label{eq:coltensor}\dzz

\medskip
Introducing the following notations
\daa
I_{(n)}(m_\pi ,b_N) &=&\int_0^\infty r^n \exp (-{r^2\over 2 b_N^2}-m_\pi r) dr
\nonumber\\
I_{(0)}(m_\pi ,b_N) &=& \sqrt{{\pi\over 2}}\; b_N
\; {\rm erfc}\bigg({m_\pi b_N\over
\sqrt{2}}\bigg) \;\exp ({m_\pi^2 b_N^2\over 2}) \nonumber\\
{\rm erfc}(z) &=& {2\over\sqrt{\pi}} \int_z^\infty \exp (-x^2) dx\nonumber\\
I_{(n+1)} &=& - {\partial\over\partial m_\pi} \; I_{(n)}
\label{eq:integrals}\dbb
for the radial integrals needed in the calculation of the Yukawa--like
pion-- and sigma--potential matrix elements, allows us to present the chiral
contributions in the following form: First, for the pion central
contribution, we write
\dyy
V^\pi_{ij}
= \frac{\Lambda^2}{\Lambda^2 - m_\pi^2}~ \frac {g^2}{4\pi}~
\frac{p_{ij}}{4m_q^2 b^3} \sqrt{\frac{2}{\pi}} \bigg\{ m_\pi^2
\bigg( p1_{ij}I_{(1)}+p3_{ij} \frac{I_{(3)}}{b^2} +p5_{ij} \frac{I_{(5)}}{b^4}
\bigg) - \Lambda^2 \bigg( m_\pi\leftrightarrow\Lambda\bigg)\bigg\}
\label{eq:pionic}\dzz
So, in the four next tables (table \ref{table:pionp} -- \ref{table:pionp5}),
we give the coefficients $p, p1, p3, p5$
respectively. Let us mention once more, that since the pion tensor
contributions involve all above integrals from $I_{(1)}$ up to
$I_{(5)}$, and gives rather negligible contributions, we do not give
the analytical expressions here.

\bigskip
Finally, table \ref{table:sigma} concerns the $\sigma$--potential,
where we give only the non--zero matrix elements. The notations are the same
as for the pion, i.e.
\dyy
V^\sigma_{ij}
= - \frac{\Lambda^2}{\Lambda^2 - m_\sigma^2}~ \frac {g^2}{4\pi}~
\frac{s_{ij}}{b^3} \sqrt{\frac{2}{\pi}} \bigg\{
\bigg( s1_{ij}I_{(1)}+s3_{ij} \frac{I_{(3)}}{b^2} +s5_{ij} \frac{I_{(5)}}{b^4}
\bigg) - \bigg( m_\sigma\leftrightarrow\Lambda\bigg)\bigg\}
\label{eq:sigisigi}\dzz

\vspace{0.5cm}
\appendix{\large\bf Appendix C}

\setcounter{equation}{0}
\renewcommand{\theequation}{C.\arabic{equation}}
\setcounter{table}{0}
\renewcommand{\thetable}{C.\arabic{table}}

\medskip
In this appendix, we present
the expectation values
of the Hamiltonian (\ref{eq:ham}) for the two states of eq.(\ref{eq:state1})
and eq.(\ref{eq:state2}):
\daa
M_1^{T=2} &=&
\langle 1\vert H^{(6q)} \vert 1\rangle \nonumber\\
&=&
6m_q + \frac {17}{4m_qb^2}
+ \frac {83a_cb}{5\sqrt{2\pi}} -4C
- \frac {77\alpha_s}{10\sqrt{2\pi}b}
+ \frac {1321\alpha_s}{360\sqrt{2\pi}m_q^2b^3} \nonumber\\
&+& V^{\pi}_C(1) + V^\sigma (1)
+ \frac {23\alpha_s}{45\sqrt{2\pi}m_q^2b^3} + V^\pi_T(1) \label{eq:mass1} \\
M_2^{T=2} &=&
\langle 2\vert H^{(6q)} \vert 2\rangle \nonumber\\
&=&
6m_q + \frac {17}{4m_qb^2}
+ \frac {256a_cb}{15\sqrt{2\pi}} -4C
- \frac {112\alpha_s}{15\sqrt{2\pi}b}
+ \frac {148\alpha_s}{45\sqrt{2\pi}m_q^2b^3} \nonumber\\
&+& V^{\pi}_C(2) + V^\sigma (2)
- \frac{\alpha_s}{15\sqrt{2\pi}m_q^2b^3} + V^\pi_T(2) \label{eq:mass2}
\dbb

One recognizes the CC--term $\simeq\frac{\alpha_s}{b}$ and the
added CE and CM--term $\simeq\frac{\alpha_s}{b^3}$. The second
$\frac{\alpha_s}{b^3}$--term comes from the tensor
part of the one--gluon exchange potential.
With the notations of appendix B, eq.(\ref{eq:pionic}) and
eq.(\ref{eq:sigisigi}), the central parts of the chiral interactions
are given in the following list

\begin{center}
\begin{tabular}{ccccccccc}
 & p & p1 & p3 & p5 & s & s1 & s3 & s5 \\ \hline
$V^\pi_C(1)$   & $\frac{1}{15}$ & 13 & $\frac{-14}{3}$ & 0 & & & & \\
$V^\sigma (1)$ & & & & & 3 & 4 & $\frac{1}{3}$ & 0 \\
$V^\pi_C(2)$   & $\frac{-1}{15}$ & 8 & 7 & 0 & & & & \\
$V^\sigma (2)$ & & & & & 3 & 4 & $\frac{1}{3}$ & 0 \\
\hline
\end{tabular}
\end{center}

The tensor contributions of the pion are given by
\dyy
V^\pi_T(i)
= {\rm pt}_{(i)} \;\frac{\Lambda^2}{\Lambda^2 - m_\pi^2}~ \frac {g^2}{4\pi}~
\frac{1}{4m_q^2 b^5} \sqrt{\frac{2}{\pi}} \bigg\{ m_\pi^2
\bigg( I_{(3)}+ \frac{3I_{(2)}}{m_\pi } + \frac{3I_{(1)}}{m_\pi^2}
\bigg) - \Lambda^2 \bigg( m_\pi\leftrightarrow\Lambda\bigg)\bigg\}
\label{eq:piontensor}\dzz
with
$$ {\rm pt}_{(1)} = \frac{-58}{45} \qquad ,\qquad
 {\rm pt}_{(2)} = \frac{1}{15} $$


\newpage
\setlength{\oddsidemargin}{ -1.5cm}
\setlength{\evensidemargin}{ -1.5cm}
\setlength{\headheight}{0cm}
\setlength{\textheight}{22cm}
\setlength{\textwidth}{18.5cm}

\setcounter{table}{0}
\renewcommand{\thetable}{\arabic{table}}

\begin{table}[h]
\begin{tabular}{|c||c|c||c|c|c|c||c|c|c|c||c|}\hline
& $b_N$ & $\Lambda$ & $m_q$ & $\alpha_s$ & $a_c$ & $C$ &
\multicolumn{4}{c||}{$N^\ast{\rm (1535)} =\lambda N_1^{(-)}+\mu N_2^{(-)}$} &
$N^{\ast\ast}$ \\ \hline
Set & [fm] & [fm$^{-1}$] & [MeV] & & [MeV/fm] & [MeV] & $\lambda$ & $\mu$ &
$N^\ast$(1535) & $N^\ast$(1650) & [MeV] \\ \hline
 I & .45 & 5.07 & 338 & .127 & 462  & 549  & .98  & .22 & 1535 & 1697 & 1678 \\
\hline
II & .47 & 7.6  & 296 & .074 & 423  & 500  & -.96 & .30 & 1535 & 1713 & 1660 \\
\hline
III& .5  & 3.55 & 274 & .334 & 654  & 676  & .998 & .06 & 1535 & 1683 & 1797 \\
\hline
 IV& .5  &  /   & 230 & .474 & 1223 & 1247 & -.86 & .51 & 1535 & 1844 & 2120 \\
\hline
 V & .6  & 10.14& 313 & .816 & 25   & -32  & .98  & .22 & 1275 & 1436 & 1292 \\
\hline
\end{tabular}
\parbox{17.5cm}{\caption{Different sets of parameters for the quark--quark
interactions of eqs.(10)--(15), fitted as described in the text.
The first 6 columns show the harmonic oscillator length $b_N$
used to describe the baryons, the cutoff $\Lambda$ parametrizing the finite
size of the pion--quark and sigma--quark vertex, the constituent quark
mass $m_q$, the strong coupling constant $\alpha_s$ of the one--gluon exchange
and the strength (slope) $a_c$ and offset $C$ of the confinement potential,
respectively.
The diagonalisation of the mass matrix in the basis of two states $N_1^{(-)}$
and $N_2^{(-)}$ gives the energies of the two negative parity resonances
N$^\ast$(1535) and N$^\ast$(1650). The next four columns give the amplitudes
and masses of these two resonances, respectively.
The last column gives for illustration the mass of the lowest two--quantum
excitation (N=2) for the parameters given, calculated in a basis of the two
most important states.
\label{table:paraset}}}
\end{table}

\begin{table}[h]
\begin{tabular}{|c||c|c||c||c|c|c|c|c|c|c|}\hline
$[f]_X$ & [51] & [51] & [42] & [42] & [42] & [411] &
[411] & [321] & [321] & [3111]\\ \hline
$(\lambda \mu)$ & (11), (30) & (11) & (11), (30)
& (11) & (30) & (11), (30)
& (11) & (11) & (11) & (00) \\ \hline
LST & 110 & 220 & 110 & 220 & 330 & 110 &
220 & 110 & 220 & 000 \\ \hline
& & [321], & [42], [321], & & & [321], &
[321], & $[42], [321]_1,$ & $[321]_1$, & \\
$[f]_{CS}$ & [321] & $[2^21^2]$ & $[31^3], [2^3],$
& [321] & $[2^3]$ & $[31^3]$  & $[2^21^2]$ &
$[321]_2, [31^3],$ & $[321]_2$, & $[41^2]$ \\
& & & $[21^4]$ & & & & & $[21^4]$ & $[2^21^2]$ & \\ \hline
& & [42], & [51], [411], & & & [411], &
[42], & [51], [411], & [42], & \\
$[f]_{ST}$ & [321] & [321] & $[33], [321],$ &
$[321]$  & [33] & [321] & [321] &
$[321]_1, [321]_2,$ & $[321]_1$, & $[31^3]$ \\
& & & $[2^21^2]$ & & & & & $[2^21^2]$ & $[321]_2$ & \\ \hline
& & [6], & [6],$[42]_1$, & & & [42], & [42], & $[42]_1,[42]_2$, &
[42], & \\
$[f]_{XS}$ & [42] & [42] & $[42]_2,[31^3]$, & [42] & [6] & $[31^3]$ &
$[31^3]$ & $[31^3]_1,[31^3]_2,$ & $[31^3]$, & $[31^3]$ \\
& & & [222] & & & & & [222] & [222] & \\ \hline
states & 2 & 2 & 10 & 1 & 1 & 4 & 2 & 5 & 3 & 1 \\ \hline
\end{tabular}
\parbox{18.5cm}{\caption{Here,
we give a complete list of all possible six--quark states with three internal
harmonic oscillator quanta N=3 and the quantum numbers J$^P$=0$^-$, T=0.
The first line defines the spatial permutational symmetry $[f]_X$, and the
second the corresponding possible Elliot symbols ($\lambda\mu $) for the
harmonic oscillator.
The third line shows the total orbital angular momentum L, the total spin S
and the total isospin T. The next three lines give the
intermediate permutational symmetry needed for an unambigious classification
of the states within a certain reduction scheme. Line 4 corresponds to the
reduction chain (29) with explicit intermediate color--spin symmetry, line 5 to
(30) with intermediate spin--isospin symmetry and line 6 to
(31) with intermediate spin--orbital symmetry. The reduction chain (31) was
used in the present calculation.
The last line gives the number of states in each column, respectively.
\label{table:states}}}
\end{table}

\begin{table}[h]
\begin{tabular}{|c||c|c||c|c||c|c||c|c|}\hline
& \multicolumn{2}{|c||}{excess energies} &
\multicolumn{2}{c||}{groundstate} & \multicolumn{2}{c||}{T=2} &
\multicolumn{2}{c|}{mixing results}\\ \hline
 & $\!\begin{array}{c} \delta M \\ (b_6=b_N)\end{array}\!$
 & $\!\begin{array}{c} \delta M_{s^6} \\ (b_6=b_N)\end{array}\!$
& $b_6$ & $M_{N=1}^{T=0}$ & $M^{T=2}_1$ & $M^{T=2}_2$ & $b_6$ & $M$
 \\ \hline
Set & $[MeV]$ & $[MeV]$ & $[fm]$ & $[MeV]$ &
$[MeV]$ & $[MeV]$ & $[fm]$ & $[MeV]$  \\ \hline
\hline
 I  & 525 & 509 & .59 & 2705 & 2564 & 2744 & .60 &  2588 \\ \hline
II  & 566 & 555 & .65 & 2680 & 2530 & 2705 & .65 &  2565 \\ \hline
III & 624 & 603 & .60 & 2898 & 2812 & 2949 & .62 &  2760 \\ \hline
IV  & 850 & 924 & .56 & 3288 & 3303 & 3333 & .59 &  3114 \\ \hline
 V  & 441 & 400 & 1.24& 2162 & 2146 & 2169 & 1.24&  2132 \\ \hline
\end{tabular}
\caption{For
different sets of parameters I to V defined in table 1, the first column
$\delta M$ gives the mass difference of the calculated lowest J$^P$=0$^-$,
T=0 state and the N(939)N$^\ast$(1535) threshold. For comparison, the second
column $\delta M_{s^6}$ gives the mass excess for six quarks, coupled to
deuteron--like quantum numbers, in one harmonic oscillator $s^6$ with respect
to the N(939)N(939) threshold at 1878 MeV. Both mass excesses in columns 1 and
2 are calculated for the oscillator length $b_6$ equal to the value $b_N$
minimizing the nucleon mass.
Columns 3 and 4 show the oscillator length $b_6$ obtained by minimizing the
single N=1, J$^P$=0$^-$, T=0, $[51]_X$, ($\lambda\mu $)=(10), $[f']_{XS}$=[42]
configuration and the corresponding mass.
Columns 5 and 6 list the masses of the two N=1, J$^P$=0$^-$, T=2 states. Here
again, the oscillator length (not given) is varied to minimize both energies
separately.
Columns 7 and 8 give the final result for the oscillator length $b_6$,
minimizing the lowest J$^P$=0$^-$, T=0 energy eigenvalue, and the
corresponding mass $M$.
This mass should be compared with the observed experimental
resonance energy of 2065 MeV.
\label{table:excess}}
\end{table}

\begin{table}[h]
\begin{tabular}{|c||c|c|c|c|c|c|c|c|c|c|c|}\hline
$\vert{\rm harm.osc.}_i\rangle $ &
$i=1$ & 2 & 3 & 4 & 5 & 6 & 7 & 8 & 9 & 10 & 11 \\ \hline\hline
$N$ & 1 & 3 & 3 & 3 & 3 & 3 & 3 & 3 & 3 & 3 & 3 \\ \hline
$[f]_X$ & [51] & [42] & [42] & [42] & [42] & [42] & [42] & [42] & [42] &
[42] & [42] \\ \hline
$(\lambda\mu)$ & (10) & (30) & (11) & (30) & (11) & (30) & (11) & (30) & (11)
& (30) & (11) \\ \hline
L, S, T & \multicolumn{11}{c|}{1, 1, 0} \\ \hline
$[f']_{XS}$ & [42] & [6] & [6] & $[2^3]$ & $[2^3]$ & $[31^3]$ & $[31^3]$ &
$[42]_1$ & $[42]_1$ & $[42]_2$ & $[42]_2$ \\ \hline
\end{tabular}
\caption{Quantum
numbers and our labelling of the states included in the configuration
mixing to calculate the lowest J$^P$=0$^-$, T=0 six--quark configuration.
The first line gives our labelling of the states.
The next line lists the
respective number of excited harmonic oscillator quanta N, and line three
represents the orbital permutational symmetry of the Young tableaux.
The fourth line shows the Elliot quantum numbers of the harmonic oscillator
($\lambda\mu $), while line five gives the total orbital angular momentum L,
the total spin S and the total isospin T of the configurations.
The last line denotes the intermediate symmetries of the Young tableaux
$[f']_{XS}$ in the orbital--spin space.
\label{table:basisstates}}
\end{table}

\setcounter{table}{0}
\renewcommand{\thetable}{B.\arabic{table}}

\begin{table}[h]
\begin{center}
\begin{tabular}{|c||c|c|c|c|c|c|c|c|c|c|c|}\hline
$a_{ij}$ & $j\! =\! 1$ & 2 & 3 & 4 & 5 & 6 & 7 & 8 & 9 & 10 & 11
\\ \hline\hline
$i=1$ &
$\frac{53}{3}$ & $\frac{5}{3\sqrt{15}}$ & $\frac{-16}{3\sqrt{
30}}$ & $\frac{5}{3\sqrt{15}}$ &
$\frac{-16}{3\sqrt{
30}}$ & $\frac{5}{3\sqrt{30}}$ &
$\frac{-8}{3\sqrt{15}}$ &
$\frac{5}{6\sqrt{15}}$ &
$\frac{-8}{3\sqrt{30}}$ &
$\frac{5}{12\sqrt{15}}$ &
$\frac{-3}{2\sqrt{30}}$ \\
\hline $2$
& & $\frac{12784}{675}$ & $\frac{4\sqrt{2}}{135}$ & 0 & 0 & 0 & 0 &
$\frac{-343}{450}$ & $\frac{-2\sqrt{2}}{45}$ & $\frac{253}{300}$ & 0 \\
\hline $3$
& & & $\frac{512}{27}$ & 0 & 0 & 0 & 0 &
$\frac{-2\sqrt{2}}{45}$ & $\frac{2}{9}$ & 0 & $\frac{1}{3}$ \\
\hline $4$
& & & & $\frac{39917}{2160}$ & $\frac{23}{540\sqrt{2}}$ &
$\frac{-9}{40\sqrt{2}}$ & $\frac{-1}{20}$ & $\frac{121}{144}$ &
$\frac{11}{180\sqrt{2}}$ & $\frac{-3}{40}$ & $\frac{-1}{30\sqrt{2}}$ \\
\hline $5$
& & & & & $\frac{2039}{108}$ & $\frac{-1}{20}$ & $\frac{1}{2\sqrt{2}}$ &
$\frac{11}{180\sqrt{2}}$ & $\frac{-1}{36}$ & $\frac{-1}{30\sqrt{2}}$ &
$\frac{1}{6}$ \\
\hline $6$
& & & & & & $\frac{20947}{1080}$ & $\frac{37}{270\sqrt{2}}$ &
$\frac{-3}{40\sqrt{2}}$ & $\frac{-1}{60}$ & $\frac{25}{18\sqrt{2}}$ &
$\frac{-1}{90}$ \\
\hline $7$
& & & & & & & $\frac{1009}{54}$ & $\frac{-1}{60}$ & $\frac{1}{6\sqrt{2}}$ &
$\frac{-1}{90}$ & $\frac{-2\sqrt{2}}{9}$ \\
\hline $8$
& & & & & & & & $\frac{203701}{10800}$ & $\frac{47}{540\sqrt{2}}$ &
$\frac{-26}{75}$ & $\frac{1}{30\sqrt{2}}$ \\
\hline $9$
& & & & & & & & & $\frac{2027}{108}$ & $\frac{1}{30\sqrt{2}}$ &
$\frac{-1}{3}$ \\
\hline $10$
& & & & & & & & & & $\frac{208441}{10800}$ & $\frac{17}{135\sqrt{2}}$ \\
\hline $11$
& & & & & & & & & & & $\frac{2021}{108}$ \\
\hline
\end{tabular}
\caption{Analytic expressions for the confinement matrix elements
\label{table:conf}}
\end{center}
\end{table}

\newcounter{ttable}
  \setcounter{ttable}{\value{table}}
  \addtocounter{ttable}{1}
  \setcounter{table}{0}
  \renewcommand{\thetable}{\mbox{B.\arabic{ttable}\alph{table}}}

\begin{table}[h]
\begin{center}
\begin{tabular}{|c||c|c|c|c|c|c|c|c|c|c|c|}\hline
$b_{ij}$ & $j=1$ & 2 & 3 & 4 & 5 & 6 & 7 & 8 & 9 & 10 & 11 \\ \hline
\hline
$i=1$
& $\frac{43}{6}$ & $\frac{-5}{6\sqrt{15}}$ & $\frac{4}{3}\sqrt{
\frac{2}{15}}$ & $\frac{-5}{6\sqrt{15}}$ &
$\frac{4}{3}\sqrt{
\frac{2}{15}}$ & $\frac{-5}{12}\sqrt{\frac{2}{15}}$ &
$\frac{4}{3\sqrt{15}}$ &
$\frac{-5}{12\sqrt{15}}$ &
$\frac{2}{3}\sqrt{\frac{2}{15}}$ &
$\frac{-5}{24\sqrt{15}}$ &
$\frac{3}{8}\sqrt{\frac{2}{15}}$ \\
\hline
$2$
& & $\frac{1528}{225}$ & $\frac{2\sqrt{2}}{45}$ & 0 & 0 & 0 & 0 &
$\frac{21}{100}$ & $\frac{\sqrt{2}}{15}$ & $\frac{-319}{600}$ & 0 \\
\hline
$3$
& & & $\frac{304}{45}$ & 0 & 0 & 0 & 0 &
$\frac{\sqrt{2}}{15}$ & $\frac{-1}{15}$ & 0 & $\frac{11}{30}$ \\
\hline
$4$
& & & & $\frac{10123}{1440}$ & $\frac{23}{360\sqrt{2}}$ &
$\frac{-7}{80\sqrt{2}}$ & $\frac{3}{40}$ & $\frac{-11}{32}$ &
$\frac{-11}{120\sqrt{2}}$ & $\frac{-13}{240}$ & $\frac{1}{20\sqrt{2}}$ \\
\hline
$5$
& & & & & $\frac{2477}{360}$ & $\frac{3}{40}$ & $\frac{-7}{20\sqrt{2}}$ &
$\frac{-11}{120\sqrt{2}}$ & $\frac{-11}{120}$ & $\frac{1}{20\sqrt{2}}$ &
$\frac{-7}{60}$ \\
\hline
$6$
& & & & & & $\frac{4853}{720}$ & $\frac{37}{180\sqrt{2}}$ &
$\frac{-13}{240\sqrt{2}}$ & $\frac{1}{40}$ & $\frac{51}{60\sqrt{2}}$ &
$\frac{1}{60}$ \\
\hline
$7$
& & & & & & & $\frac{1243}{180}$ & $\frac{1}{40}$ & $\frac{-7}{60\sqrt{2}}$ &
$\frac{1}{60}$ & $\frac{22\sqrt{2}}{135}$ \\
\hline
$8$
& & & & & & & & $\frac{49859}{7200}$ & $\frac{47}{360\sqrt{2}}$ &
$\frac{8}{25}$ & $\frac{-1}{20\sqrt{2}}$ \\
\hline
$9$
& & & & & & & & & $\frac{2489}{360}$ & $\frac{-1}{20\sqrt{2}}$ &
$\frac{59}{90}$ \\
\hline
$10$
& & & & & & & & & & $\frac{48719}{7200}$ & $\frac{17}{90\sqrt{2}}$ \\
\hline
$11$
& & & & & & & & & & & $\frac{2483}{360}$ \\
\hline
\end{tabular}
\caption{Analytic expressions for the color coulomb  matrix elements
\label{table:coco}}
\end{center}
\end{table}

\begin{table}[h]
\begin{center}
\begin{tabular}{|c||c|c|c|c|c|c|c|c|c|c|c|}\hline
$ce_{ij}$ & $j=1$ & 2 & 3 & 4 & 5 & 6 & 7 & 8 & 9 & 10 & 11 \\ \hline\hline
$i=1$ & $\frac{99}{72}$ & $\frac{-5}{8\sqrt{15}}$ & $\frac{2}{\sqrt{30}}$ &
  $\frac{-5}{8\sqrt{15}}$ & $\frac{2}{\sqrt{30}}$ &
  $\frac{-5}{8\sqrt{30}}$ & $\frac{1}{\sqrt{15}}$ &
  $\frac{-5}{16\sqrt{15}}$ & $\frac{1}{\sqrt{30}}$ &
  $\frac{-5}{32\sqrt{15}}$ & $\frac{9}{16\sqrt{30}}$ \\ \hline
$2$ & & $\frac{58}{45}$ & $\frac{-1}{9\sqrt{2}}$ & 0 & 0 & 0 & 0 &
  $\frac{7}{240}$ & $\frac{\sqrt{2}}{12}$ & $\frac{-77}{160}$ & 0 \\ \hline
$3$ & & & $\frac{56}{45}$ & 0 & 0 & 0 & 0 & $\frac{\sqrt{2}}{12}$ &
  $\frac{-1}{60}$ & 0 & $\frac{-17}{40}$ \\ \hline
$4$ & & & & $\frac{1699}{1152}$ & $\frac{-23}{288\sqrt{2}}$ &
  $\frac{-21}{64\sqrt{2}}$ & $\frac{3}{32}$ & $\frac{-77}{384}$ &
  $\frac{-11}{96\sqrt{2}}$ & $\frac{-7}{64}$ & $\frac{1}{16\sqrt{2}}$ \\ \hline
$5$ & & & & & $\frac{2017}{1440}$ & $\frac{3}{32}$ & $\frac{-27}{80\sqrt{2}}$ &
  $\frac{-11}{96\sqrt{2}}$ & $\frac{-71}{480}$ & $\frac{1}{16\sqrt{2}}$ &
  $\frac{-9}{80}$ \\ \hline
$6$ & & & & & & $\frac{797}{576}$ & $\frac{-37}{144\sqrt{2}}$ &
  $\frac{-7}{64\sqrt{2}}$ & $\frac{1}{32}$ &
  $\frac{35}{48\sqrt{2}}$ & $\frac{1}{48}$ \\ \hline
$7$ & & & & & & & $\frac{983}{720}$ & $\frac{1}{32}$ & $\frac{-9}{80\sqrt{2}}$
&
  $\frac{1}{48}$ & $\frac{19}{30\sqrt{2}}$ \\ \hline
$8$ & & & & & & & & $\frac{8411}{5760}$ & $\frac{-47}{288\sqrt{2}}$ &
  $\frac{-3}{16}$ & $\frac{-1}{16\sqrt{2}}$ \\ \hline
$9$ & & & & & & & & & $\frac{2029}{1440}$ &
  $\frac{-1}{16\sqrt{2}}$ & $\frac{13}{40}$ \\ \hline
$10$ & & & & & & & & & & $\frac{7991}{5760}$ & $\frac{-17}{72\sqrt{2}}$\\
\hline
$11$ & & & & & & & & & & & $\frac{1963}{1440}$ \\ \hline
\end{tabular}
\caption{Analytic expressions for the color electric
matrix elements ($\delta$--force)
\label{table:coel}}
\end{center}
\end{table}

\begin{table}[h]
\begin{center}
\begin{tabular}{|c||c|c|c|c|c|c|c|c|c|c|c|}\hline
$cm_{ij}$ & $j=1$ & 2 & 3 & 4 & 5 & 6 & 7 & 8 & 9 & 10 & 11 \\ \hline\hline
$i=1$ & $\frac{5}{36}$ & $\frac{-5}{12\sqrt{15}}$ & $\frac{4}{3\sqrt{30}}$ &
  $\frac{-5}{12\sqrt{15}}$ & $\frac{4}{3\sqrt{30}}$ &
  $\frac{5}{4\sqrt{30}}$ & $\frac{-2}{\sqrt{15}}$ &
  $\frac{-25}{24\sqrt{15}}$ & $\frac{4}{3\sqrt{30}}$ &
  $\frac{85}{144\sqrt{15}}$ & $\frac{-53}{72\sqrt{30}}$ \\ \hline
$2$ & & $\frac{116}{135}$ & $\frac{-2}{27\sqrt{2}}$ & 0 & 0 & 0 & 0 &
  $\frac{7}{360}$ & $\frac{\sqrt{2}}{18}$ & $\frac{-77}{240}$ & 0 \\ \hline
$3$ & & & $\frac{112}{135}$ & 0 & 0 & 0 & 0 & $\frac{\sqrt{2}}{18}$ &
  $\frac{-1}{90}$ & 0 & $\frac{-17}{60}$ \\ \hline
$4$ & & & & $\frac{-1807}{1728}$ & $\frac{49}{432\sqrt{2}}$ &
  $\frac{21}{32\sqrt{2}}$ & $\frac{-3}{16}$ & $\frac{-77}{576}$ &
  $\frac{-11}{144\sqrt{2}}$ & $\frac{-7}{96}$ & $\frac{1}{24\sqrt{2}}$ \\
\hline
$5$ & & & & & $\frac{-142}{4320}$ & $\frac{-3}{16}$ & $\frac{27}{40\sqrt{2}}$ &
  $\frac{-11}{144\sqrt{2}}$ & $\frac{-71}{720}$ & $\frac{1}{24\sqrt{2}}$ &
  $\frac{-9}{40}$ \\ \hline
$6$ & & & & & & $\frac{-13}{288}$ & $\frac{-11}{72\sqrt{2}}$ &
  $\frac{7}{32\sqrt{2}}$ & $\frac{-1}{16}$ &
  $\frac{-35}{24\sqrt{2}}$ & $\frac{-1}{24}$ \\ \hline
$7$ & & & & & & & $\frac{723}{720}$ & $\frac{-1}{16}$ & $\frac{9}{40\sqrt{2}}$
&
  $\frac{-1}{24}$ & $\frac{-19}{15\sqrt{2}}$ \\ \hline
$8$ & & & & & & & & $\frac{3731}{8640}$ & $\frac{-23}{432\sqrt{2}}$ &
  $\frac{109}{144}$ & $\frac{-7}{72\sqrt{2}}$ \\ \hline
$9$ & & & & & & & & & $\frac{853}{2160}$ &
  $\frac{-7}{72\sqrt{2}}$ & $\frac{37}{180}$ \\ \hline
$10$ & & & & & & & & & & $\frac{3331}{8640}$ & $\frac{-13}{144\sqrt{2}}$
\\ \hline
$11$ & & & & & & & & & & & $\frac{887}{2160}$ \\ \hline
\end{tabular}
\caption{Analytic expressions for the color magnetic
matrix elements ($\delta$--force)
\label{table:coma}}
\end{center}
\end{table}

\begin{table}[h]
\begin{center}
\begin{tabular}{|c||c|c|c|c|c|c|c|c|c|c|c|}\hline
$ct_{ij}$ & $j=1$ & 2 & 3 & 4 & 5 & 6 & 7 & 8 & 9 & 10 & 11 \\ \hline\hline
$i=1$ & $\frac{1}{12}$ & $\frac{4}{15\sqrt{15}}$ & $\frac{1}{6\sqrt{30}}$ &
  $\frac{5}{12\sqrt{15}}$ & $\frac{-7}{12\sqrt{30}}$ & 0 & 0 &
  $\frac{-29}{90\sqrt{15}}$ & $\frac{-1}{4\sqrt{30}}$ &
  $\frac{257}{360\sqrt{15}}$ & $\frac{29}{36\sqrt{30}}$ \\ \hline
$2$ & & $\frac{74}{3375}$ & $\frac{16}{1350\sqrt{2}}$ & 0 & 0 & 0 & 0 &
  $\frac{133}{3375}$ & $\frac{-7\sqrt{2}}{1800}$ & $\frac{1}{125}$ &
  $\frac{-1}{200\sqrt{2}}$ \\ \hline
$3$ & & & $\frac{2}{135}$ & 0 & 0 & 0 & 0 & $\frac{-7\sqrt{2}}{1800}$ &
  $\frac{-1}{45}$ & $\frac{-1}{200\sqrt{2}}$ & 0 \\ \hline
$4$ & & & & $\frac{-1}{108}$ & $\frac{1}{54\sqrt{2}}$ &
  $\frac{3}{5\sqrt{2}}$ & $\frac{-1}{20}$ & $\frac{-1}{90}$ &
  $\frac{-1}{360\sqrt{2}}$ & $\frac{1}{120}$ & $\frac{-1}{144\sqrt{2}}$
  \\ \hline
$5$ & & & & & $\frac{-11}{108}$ & $\frac{-1}{20}$ & $\frac{1}{5\sqrt{2}}$ &
  $\frac{-1}{360\sqrt{2}}$ & $\frac{7}{90}$ & $\frac{-1}{144\sqrt{2}}$ &
  $\frac{-1}{120}$ \\ \hline
$6$ & & & & & & 0 & 0 &
  $\frac{1}{5\sqrt{2}}$ & $\frac{-1}{60}$ & 0 & 0 \\ \hline
$7$ & & & & & & & 0 & $\frac{-1}{60}$ & $\frac{4}{45\sqrt{2}}$ & 0 & 0 \\
\hline
$8$ & & & & & & & & $\frac{-2717}{40500}$ & $\frac{\sqrt{2}}{675}$ &
  $\frac{-526}{3375}$ & $\frac{109}{3600\sqrt{2}}$ \\ \hline
$9$ & & & & & & & & & $\frac{2771}{12960}$ &
  $\frac{109}{3600\sqrt{2}}$ & $\frac{37}{1620}$ \\ \hline
$10$ & & & & & & & & & & $\frac{-3127}{10800}$ &
$\frac{11}{5400\sqrt{2}}$\\ \hline
$11$ & & & & & & & & & & & $\frac{-103}{4320}$ \\ \hline
\end{tabular}
\caption{Analytic expressions for the gluon tensor matrix elements
\label{table:cot}}
\end{center}
\end{table}

\setcounter{table}{\value{ttable}}
\renewcommand{\thetable}{B.\arabic{table}}

\newcounter{tatable}
  \setcounter{tatable}{\value{table}}
  \addtocounter{tatable}{1}
  \setcounter{table}{0}
  \renewcommand{\thetable}{\mbox{B.\arabic{tatable}\alph{table}}}

\begin{table}[h]
\begin{center}
\begin{tabular}{|c||c|c|c|c|c|c|c|c|c|c|c|}\hline
$p_{ij}$ &
$j=1$ & 2 & 3 & 4 & 5 & 6 & 7 & 8 & 9 & 10 & 11 \\ \hline\hline
i=1 & $\frac{1}{3}$ & $\frac{-10}{3\sqrt{15}}$ & $\frac{16}{3}\sqrt{
\frac{2}{15}}$ & $\frac{-10}{3\sqrt{15}}$ &
$\frac{16}{3}\sqrt{
\frac{2}{15}}$ & $-5\sqrt{\frac{2}{15}}$ &
$\frac{16}{\sqrt{15}}$ &
$\frac{25}{3\sqrt{15}}$ &
$\frac{-4}{3}\sqrt{\frac{2}{15}}$ &
$\frac{-55}{6\sqrt{15}}$ &
$\frac{47}{6}\sqrt{\frac{2}{15}}$ \\
\hline $2$
& & $\frac{-1}{3}$ & $\frac{\sqrt{2}}{12}$ & 0 & 0 & 0 & 0 &
$\frac{7}{9}$ & $\frac{2\sqrt{2}}{9}$ & $\frac{-1}{6}$ & 0 \\
\hline $3$
& & & $\frac{-1}{15}$ & 0 & 0 & 0 & 0 &
$\frac{2\sqrt{2}}{9}$ & $\frac{-2}{45}$ & 0 & $\frac{-1}{15}$ \\
\hline $4$
& & & & $\frac{1}{8}$ & $\frac{-3}{4\sqrt{2}}$ &
$\frac{-1}{4\sqrt{2}}$ & $\frac{3}{2}$ & $\frac{-1}{72}$ &
$\frac{-11}{36\sqrt{2}}$ & $\frac{-1}{12}$ & $\frac{1}{6\sqrt{2}}$ \\
\hline $5$
& & & & & $\frac{1}{10}$ & $\frac{3}{2}$ & $\frac{-1}{10\sqrt{2}}$ &
$\frac{-11}{36\sqrt{2}}$ & $\frac{-1}{90}$ & $\frac{1}{6\sqrt{2}}$ &
$\frac{-1}{5}$ \\
\hline $6$
& & & & & & $\frac{1}{4}$ & $\frac{-1}{3\sqrt{2}}$ &
$\frac{-1}{4\sqrt{2}}$ & $\frac{1}{2}$ & $\frac{1}{3\sqrt{2}}$ &
$\frac{1}{3}$ \\
\hline $7$
& & & & & & & $\frac{1}{15}$ & $\frac{1}{2}$ & $\frac{-1}{5\sqrt{2}}$ &
$\frac{1}{3}$ & $\frac{4}{15\sqrt{2}}$ \\
\hline $8$
& & & & & & & & $\frac{1}{72}$ & $\frac{-11}{36\sqrt{2}}$ &
$\frac{1}{6}$ & $\frac{1}{6\sqrt{2}}$ \\
\hline $9$
& & & & & & & & & $\frac{1}{90}$ & $\frac{1}{6\sqrt{2}}$ &
$\frac{1}{15}$ \\
\hline $10$
& & & & & & & & & & $\frac{1}{72}$ & $\frac{-2}{9\sqrt{2}}$ \\
\hline $11$
& & & & & & & & & & & $\frac{1}{45}$ \\
\hline
\end{tabular}
\parbox{16cm}{
\caption{Analytic expressions for the pion exchange matrix elements
(central part)
\label{table:pionp}}}
\end{center}
\end{table}

\begin{table}[h]
\begin{center}
\begin{tabular}{|c||c|c|c|c|c|c|c|c|c|c|c|}\hline
p1$_{ij}$ &
$j=1$ & 2 & 3 & 4 & 5 & 6 & 7 & 8 & 9 & 10 & 11 \\ \hline\hline
$i=1$ & 5 & 1 & 1 &
1 & 1 & 1 & 1 & 1 & 1 & 1 & 1 \\
\hline $2$
& & $\frac{29}{5}$ &
1 &0&0&0&0&
$\frac{1}{10}$ &
1 & $\frac{77}{10}$ & 0 \\
\hline $3$
& & & 28 &0&0&0&0& 1 & 1 & 0 & 17 \\
\hline $4$
& & & & $\frac{69}{2}$ & 1 &
$\frac{21}{2}$ & 1 &
$\frac{77}{2}$ & 1 &
$\frac{7}{2}$ & 1 \\
\hline $5$
& & & & & $\frac{87}{2}$ & 1 & 27 & 1 &
 $\frac{71}{2}$ &
1 & $\frac{3}{2}$ \\
\hline $6$
& & & & & & $\frac{33}{2}$ &
1 & $\frac{7}{2}$ &
1 & $\frac{35}{2}$ & 1 \\
\hline $7$
& & & & & & & $\frac{113}{2}$ &
1 & $\frac{9}{2}$ & 1 & 19 \\
\hline $8$
& & & & & & & & $\frac{1277}{10}$ &
1 & $\frac{61}{10}$ & 1 \\
\hline $9$
& & & & & & & & & $\frac{331}{2}$ & 1 & 14 \\
\hline $10$
& & & & & & & & & & $\frac{1127}{10}$ & 1 \\
\hline $11$
& & & & & & & & & & & $\frac{259}{4}$ \\
\hline
\end{tabular}
\caption{p1 for the analytic expressions for the pion exchange matrix elements
\label{table:pionp1}}
\end{center}
\end{table}

\begin{table}[h]
\begin{center}
\begin{tabular}{|c||c|c|c|c|c|c|c|c|c|c|c|}\hline
p3$_{ij}$ &
$j=1$ & 2 & 3 & 4 & 5 & 6 & 7 & 8 & 9 & 10 & 11 \\ \hline\hline
$i=1$ & -2 & $\frac{-1}{3}$ & $\frac{-1}{3}$ &
$\frac{-1}{3}$ & $\frac{-1}{3}$ &
$\frac{-1}{3}$ & $\frac{-1}{3}$ &
$\frac{-1}{3}$ & $\frac{-1}{3}$ &
$\frac{-1}{3}$ & $\frac{-1}{3}$ \\
\hline $2$
& & $\frac{-11}{3}$ &
$\frac{-2}{3}$ &0&0&0&0&
$\frac{1}{3}$ &
$\frac{-2}{3}$ & 7 & 0 \\
\hline $3$
& & & $\frac{-53}{3}$ &0&0&0&0&
$\frac{-2}{3}$ &
$\frac{4}{3}$ & 0 & 16 \\
\hline $4$
& & & & -9 &
$\frac{-2}{3}$ &
$\frac{-11}{3}$ &
$\frac{-2}{3}$ & 61 &
$\frac{-2}{3}$ &
$\frac{7}{3}$  &
$\frac{-2}{3}$ \\
\hline $5$
& & & & & -7 &
$\frac{-2}{3}$ & $\frac{-2}{3}$ &
$\frac{-2}{3}$ & $\frac{97}{3}$ &
$\frac{-2}{3}$ & $\frac{13}{9}$ \\
\hline $6$
& & & & & & $\frac{-49}{9}$ &
$\frac{-2}{3}$ & $\frac{-11}{9}$ &
$\frac{-2}{3}$ & $\frac{1}{3}$ &
$\frac{-2}{3}$ \\
\hline $7$
& & & & & & & $\frac{-64}{3}$ &
$\frac{-2}{3}$ & $\frac{-1}{9}$ &
$\frac{-2}{3}$ & $\frac{-13}{6}$ \\
\hline $8$
& & & & & & & & $\frac{41}{3}$ &
$\frac{-2}{3}$ & $\frac{19}{3}$ &
$\frac{-2}{3}$ \\
\hline $9$
& & & & & & & & & $\frac{107}{3}$ &
$\frac{-2}{3}$ & $\frac{41}{3}$ \\
\hline $10$
& & & & & & & & & & $\frac{-25}{3}$ &
$\frac{-2}{3}$ \\
\hline $11$
& & & & & & & & & & & $\frac{-11}{3}$ \\
\hline
\end{tabular}
\caption{p3 for the analytic expressions for the pion exchange matrix elements
\label{table:pionp3}}
\end{center}
\end{table}

\begin{table}[h]
\begin{center}
\begin{tabular}{|c||c|c|c|c|c|c|c|c|c|c|c|}\hline
p5$_{ij}$ &
$j=1$ & 2 & 3 & 4 & 5 & 6 & 7 & 8 & 9 & 10 & 11 \\ \hline\hline
$i=1$ & 0 & 0 & 0 & 0 & 0 & 0 & 0 & 0 & 0 & 0 & 0 \\
\hline $2$
& & $\frac{11}{75}$ &
$\frac{1}{15}$ &0&0&0&0&
$\frac{-11}{150}$ &
$\frac{1}{15}$ & $\frac{11}{50}$ & 0 \\
\hline $3$
& & & $\frac{2}{3}$ &0&0&0&0&
$\frac{1}{15}$ &
$\frac{-1}{3}$ & 0 & 1 \\
\hline $4$
& & & & $\frac{11}{10}$ &
$\frac{1}{15}$ &
$\frac{11}{10}$ &
$\frac{1}{15}$ &
$\frac{-11}{6}$ &
$\frac{1}{15}$ &
$\frac{11}{30}$ &
$\frac{1}{15}$ \\
\hline $5$
& & & & & $\frac{1}{2}$ &
$\frac{1}{15}$ & 1 &
$\frac{1}{15}$ & $\frac{19}{6}$ &
$\frac{1}{15}$ & $\frac{1}{18}$ \\
\hline $6$
& & & & & & $\frac{11}{18}$ &
$\frac{1}{15}$ & $\frac{11}{30}$ &
$\frac{1}{15}$ & $\frac{11}{30}$ &
$\frac{1}{15}$ \\
\hline $7$
& & & & & & & $\frac{17}{6}$ &
$\frac{1}{15}$ & $\frac{1}{6}$ &
$\frac{1}{15}$ & $\frac{7}{6}$ \\
\hline $8$
& & & & & & & & $\frac{473}{150}$ &
$\frac{1}{15}$ & $\frac{-11}{150}$ &
$\frac{1}{15}$ \\
\hline $9$
& & & & & & & & & $\frac{-1}{6}$ &
$\frac{1}{15}$ & $\frac{1}{3}$ \\
\hline $10$
& & & & & & & & & & $\frac{803}{150}$ &
$\frac{1}{15}$ \\
\hline $11$
& & & & & & & & & & & $\frac{41}{12}$ \\
\hline
\end{tabular}
\caption{p5 for the analytic expressions for the pion exchange matrix elements
\label{table:pionp5}}
\end{center}
\end{table}

\setcounter{table}{\value{tatable}}
\renewcommand{\thetable}{B.\arabic{table}}

\begin{table}[h]
\begin{center}
\begin{tabular}{|c||c|c|c|c|}\hline
$(ij)$ & $s_{ij}$ & $s1_{ij}$ & $s3_{ij}$ & $s5_{ij}$ \\ \hline\hline
$(11)$ & 1 & 12 & 1 & 0 \\ \hline
$(22),(44),(66),(88),(10,10)$ & $\frac{1}{3}$ & 29 &
$\frac{5}{3}$ & $\frac{11}{15}$ \\ \hline
$(33),(55),(77),(99),(11,11)$ & $\frac{1}{3}$ & 28 &
$\frac{7}{3}$ & $\frac{2}{3}$ \\ \hline
$(23),(45),(67),(89),(10,11)$ & $\frac{-5}{6\sqrt{2}}$ & 1 &
$\frac{-2}{3}$ & $\frac{1}{15}$ \\ \hline
\end{tabular}
\caption{Analytic expressions for the non-zero sigma exchange matrix elements
\label{table:sigma}}
\end{center}
\end{table}

\end{document}